# Computational Investigation on the formation of liquid-fueled oblique detonation waves

Wenhao Wang[a,b,c], Zongmin Hu[a,c,*], and Peng Zhang[b,*]

[a] *State Key Laboratory of High-Temperature Gas Dynamics (LHD), Institute of Mechanics, Chinese Academy of Sciences, Beijing 100190, China*

[b] *Department of Mechanical Engineering, City University of Hong Kong, Kowloon Tong, Kowloon 999077, Hong Kong*

[c] *School of Engineering Sciences, University of Chinese Academy of Sciences, Beijing 100049, China*

---

**Abstract**

Utilizing a two-phase supersonic chemically reacting flow solver with the Eulerian-Lagrangian method implemented in OpenFOAM, this study computationally investigates the formation of liquid-fueled oblique detonation waves (ODWs) within a pre-injection oblique detonation wave engine operating at an altitude of 30 km and a velocity of Mach 9. The inflow undergoes two-stage 12.5° compression, followed by uniform mixing with randomly distributed n-heptane droplets before entering the combustor. The study examines the effects of droplet breakup models, gas-liquid ratios, and on-wedge strips on the ODW formation. Results indicate that under the pure-droplet condition, the ODW fails to form within the combustor, irrespective of the breakup models used. However, increasing the proportion of n-heptane vapor in the fuel/air mixture facilitates the ODW formation, because the n-heptane vapor rapidly participates in the gaseous reactions, producing heat and accelerating the transition from low- to intermediate-temperature chemistry. Additionally, the presence of on-wedge strips enhances ODW formation by inducing a bow shock wave within the combustor, which significantly increases the temperature, directly triggering intermediate-temperature chemistry and subsequent heat-release reactions, thereby facilitating the formation of ODW.

---

*Corresponding author.

Zongmin Hu: huzm@imech.ac.cn

Peng Zhang: penzhang@cityu.edu.hk

## Nomenclature

| Symbol | Description |
|---|---|
| $A_p$ | surface area of droplet ($m^2$) |
| $B_M$ | mass transfer number (-) |
| $D$ | density-weighted diffusion coefficient ($m^2$/s) |
| $\overline{D}_f$ | average binary diffusion coefficient ($m^2$/s) |
| $D_p$ | Droplet diameter (m) |
| $D_s$ | child droplet's diameter (m) |
| $E$ | total energy (J) |
| $e$ | internal energy (J) |
| $n(t)$ | droplet number (-) |
| $\boldsymbol{q}$ | diffusive heat flux vector (W/$m^2$) |
| $t_b$ | breakup completion time (s) |
| $t_i$ | initial breakup time (s) |
| $\dot{\omega}_i$ | net production of $i$-th species (kg/$m^3$/s) |

***Greek letter***

| Symbol | Description |
|---|---|
| $\rho$ | Density (kg/$m^3$) |
| $\mu$ | dynamic viscosity (N·s/$m^2$) |
| $\nu$ | kinematic viscosity coefficient ($m^2$/s) |
| $\boldsymbol{\tau}$ | deviatoric stress tensor (Pa) |

**Abbreviation**

| Symbol | Description |
|---|---|
| $c_p$ | heat capacity of droplet (J/K) |
| $C_p$ | heat capacity at a constant pressure of gas phase (J/K) |
| $C_d$ | drag coefficient (-) |
| $D_p$ | droplet diameter (m) |
| $\boldsymbol{F}_p$ | droplet force (N) |
| $h_c$ | convective heat transfer coefficient (W/$m^2$) |
| $MW_i$ | molecular weight of species $i$ (kg/mol) |
| $\dot{m}_p$ | droplet evaporation rate (kg/s) |
| $nq$ | number of chemical reactions (-) |
| $ns$ | number of species (-) |
| $N_p$ | particle number in a cell (-) |
| $Nu$ | Nusselt number: $Nu = hL/k$ |
| $Pr$ | Prandtl number: $Pr = c_p\mu/k$ |
| $p$ | pressure (Pa) |
| $\dot{Q}_c$ | convective heat transport rate (J/s) |
| $\dot{Q}_l$ | latent heat transport rate (J/s) |
| $R$ | specific gas constant (J/kg/K) |
| $R_u$ | universal gas constant (J/mol/K) |
| $Sh$ | Sherwood number: $Sh = hL/\overline{D}_f$ |
| $Sc$ | Schmidt number: $Sc = \nu/\overline{D}_f$ |
| $T$ | gas temperature (K) |
| $T_p$ | droplet temperature (K) |
| $\boldsymbol{u}$ | gas phase velocity vector (m/s) |
| $\boldsymbol{u}_p$ | droplet velocity vector (m/s) |
| $V_c$ | calculation cell volume ($m^3$) |
| $We_p$ | droplet Weber number (-) |
| $Y_i$ | mass fraction of species $i$ (-) |
| $\rho_p$ | droplet density (kg/$m^3$) |
| $\lambda_{KH}$ | Kelvin–Helmholtz wavelength (m) |
| $\lambda_{RT}$ | Rayleigh–Taylor wavelength (m) |
| $\sigma_p$ | droplet surface tension (N/m) |
| $\tau_{RT}$ | RT breakup characteristic time (s) |
| $\tau_{KH}$ | KH breakup characteristic time (s) |
| BSW | bow shock wave |
| CW | compression wave |
| KET | keto-heptyl peroxide |
| KHI | Kelvin–Helmholtz instability |
| MS | Mach stem |
| OSW | oblique shock wave |
| ODW | oblique detonation wave |
| ODWE | oblique detonation wave engine |
| PSI-CELL | particle-source-in-cell |
| RSW | reflected shock wave |
| RTI | Rayleigh–Taylor instability |
| SSW | separation shock wave |
| SODW | secondary oblique detonation wave |

## 1. Introduction

Oblique detonation wave (ODW) is a shock-induced combustion wave phenomenon, in which the flame front is tightly coupled with the shock wave and confined to a thin region. An ODW can be stabilized by a wedge in a hypersonic flow of a fuel-air combustible mixture [1-3]. The concept of using stabilized ODWs for hypersonic propulsion was first proposed by Dunlap et al. [1] and has advanced the development of oblique detonation wave engines (ODWE), gaining increasing interest in recent years [4-6]. As an extension of the scramjet (supersonic combustion ramjet) at high Mach numbers, ODWE has emerged as one of the viable options for hypersonic propulsion systems owing to its exceptional theoretical performance and short combustor length [7, 8]. Although liquid hydrocarbon fuels stand out as the primary fuel options for ODWEs due to their high energy density and ease of storage, the long ignition delay time of hydrocarbon fuels presents a significant challenge for initiating and stabilizing the ODW in the combustor, and the liquid fuel atomization and evaporation make this challenge more serious.

In the past few decades, extensive computational and experimental research has been conducted on the initiation and stabilization of gaseous ODWs over wedges. The initiation and structure of wedge-induced hydrogen/air ODWs were first numerically studied by Li et al. [9]. They discovered a multidimensional detonation structure comprising a nonreactive oblique shock, an induction zone, a series of deflagration waves, and a reactive shock, being referred to as the saltation ODW. Viguier et al. [10] experimentally confirmed the existence of this structure in a shock tube. Figueria et al. [11] and Papalexandris et al. [12] numerically identified another initiation structure, where the nonreactive shock and reactive shock are connected by a smoothly curved shock wave, being referred to as the smoothness ODW. The initiation structures are influenced by various operating conditions such as the inflow pressure, inflow Mach number, wedge angle, and fuel-air ratio [13-17]. Similar to the oblique shock wave (OSW), the ODW has a standing window of wedge angles within which it can be attached and stabilized on the wedge [18]. The instability of ODWs can be influenced by the fuel combustion characteristics such as the reaction activation energy and heat release [19, 20].

The understanding of wedge-induced ODWs has prompted investigations into ODWs in real combustors of ODWEs. Two configurations of ODWEs have been proposed and are distinguished by the different locations of fuel injection. The first is the pre-injection ODWE, where fuel is injected and mixed in the inlet, and then the fuel/air mixture is ignited in the form of ODWs in the combustor. This configuration is also the focus of the present study. The second is known as the shock-induced combustion ramjet, where fuel is injected into the isolator, like that adopted by the scramjet. Sislian et al. conducted extensive research on both configurations [21-24] for hydrogen-fueled ODWEs.

Zhang et al. [25] designed a typical hydrogen-fueled pre-injection ODWE

combustor and numerically analyzed the wave structure, stabilization characteristics, and potential thrust under uniform premixed inflow conditions. Two stabilized detonation modes, namely the ODW mode and the normal detonation wave (NDW) mode, exist in their combustor and are influenced by the position of the designed reflecting point of the ODW. They also numerically investigated the hydrogen-air mixing using strut injectors in the inlet and the subsequent combustion in the combustor [6], where the stable ODW combustion mode proved the concept of the pre-injection ODWE. Subsequently, Zhang et al. [26] successfully conducted the first large-scale hydrogen-fueled ODWE experiment in a hypersonic high-enthalpy shock tunnel, thereby demonstrating the feasibility of gaseous oblique detonation propulsion.

In liquid-fueled ODWEs, the fuel atomization and evaporation make the ignition process more difficult. Additionally, the gas-liquid two-phase interactions complicate the structure and stabilization of the ODW. Ren et al. [27] conducted the first numerical study on wedge-induced ODWs in kerosene/air mixtures using a hybrid Eulerian-Lagrangian method and Particle-source-in-cell (PSI-CELL) method to account for the two-phase coupling. Successfully stabilizing the ODW on the wedge, they investigated the influence of droplet mass flow rate and droplet diameter on the initiation mode of ODW. Ren et al. [28] also performed a parametric study on the effects of fuel-air ratio and inflow Mach numbers on ODW initiation. In addition, they examined the effect of oscillating pressure on the stabilization of ODWs and found that the ODWs are resilient to pressure disturbances [29].

Zhang et al. [30] developed a two-phase supersonic reactive flow solver named RYrhoCentralFoam based on the OpenFOAM platform. Using this solver, Guo et al. [31] investigated the autoignition and transition mode of n-heptane droplet/vapor/air mixtures behind an oblique shock wave. They utilized a closed reactor with a size of $1\times1\times1$ mm$^3$ to simulate the ignition process, given the resemblance between detonation combustion and constant-volume combustion. Teng et al. [32] conducted numerical simulations of ODWs in partially pre-vaporized n-heptane droplets/air mixture and observed that the initiation lengths of oblique detonation waves nonmonotonically vary with the droplet diameter. There are few numerical and experimental investigations of liquid-fueled ODWs in real combustors. Han et al. [33] conducted the first liquid-fueled ODWE experiment in a hypersonic shock tunnel and found that the formation of ODWs was enhanced by employing an on-wedge strip.

Despite the aforementioned experimental achievement of liquid-phase oblique detonation within combustors, the formation conditions for liquid-fueled detonation waves are still insufficiently understood due to the scarce experimental data. Therefore, this study aims to numerically investigate n-heptane-fueled ODWs in a typical pre-injection ODWE combustor, comprehensively examining the impact of gas-liquid ratios, droplet breakup models, and on-wedge strips on the ignition and combustion modes of the ODWE combustor. The computational methodology and case settings are presented in Section 2. Numerical validation is addressed in Section 3.1, followed by results within the combustor in Section 3.2. Section 3.2.1 introduces the flow structure, while Sections 3.2.2, 3.2.3, and 3.2.4 discuss the influence of droplet breakup models, gas-liquid ratios, and on-wedge strips, respectively. Conclusions are drawn, and future works are suggested in Section 4.

## 2. Computational methodology

### *2.1. Governing Equations and sub-models*

The present study adopts an Eulerian-Lagrangian approach to simulate the two-phase compressible chemically reacting flow. In the Eulerian approach to simulate the gas flow, the governing equations for the compressible flow are given by

$$\frac{\partial \rho}{\partial t} + \nabla \cdot (\rho \boldsymbol{u}) = S_m, \tag{1}$$

$$\frac{\partial (\rho \boldsymbol{u})}{\partial t} + \nabla \cdot [\boldsymbol{u}(\rho \boldsymbol{u})] + \nabla p - \nabla \cdot \boldsymbol{\tau} = \boldsymbol{S_F}, \tag{2}$$

$$\frac{\partial (\rho E)}{\partial t} + \nabla \cdot [\boldsymbol{u}(\rho E)] + \nabla \cdot (\boldsymbol{u} p) + \nabla \cdot \boldsymbol{q} - \nabla \cdot (\boldsymbol{\tau} \cdot \boldsymbol{u}) = S_e, \tag{3}$$

$$\frac{\partial (\rho Y_i)}{\partial t} + \nabla \cdot [\boldsymbol{u}(\rho Y_i)] + \nabla \cdot [-D\nabla(\rho Y_i)] = \dot{\omega}_i + S_{s,i}, (i = 1,2, \dots, ns-1). \tag{4}$$

In Equ. (1) - (4), the variables $\rho$, $\boldsymbol{u}$, and $p$ are the gas density, velocity, and pressure, respectively; $\boldsymbol{\tau}$ is deviatoric stress tensor expressed as $\boldsymbol{\tau} = \mu \Big[\nabla \boldsymbol{u} +$ $(\nabla \boldsymbol{u})^T - \frac{2}{3}(\nabla \cdot \boldsymbol{u})\mathbf{I}\Big]$, in which $\mu$ is the dynamic viscosity calculated by Sutherland's law; $E = e + \frac{1}{2}|\boldsymbol{u}|^2$ is the total energy, in which $e$ is the internal energy; $\boldsymbol{q}$ is the diffusive heat flux calculated by Fourier's law as $\boldsymbol{q} = -k\nabla T$ with $k$ being the thermal conductivity; $ns$ is the number of species, and $Y_1, \dots, Y_{ns-1}$ are the mass fractions of each species; $D$ is the density-weighted diffusion coefficient which can be calculated with the unity Lewis number assumption as $D = k/\rho C_p$. Here, $C_p$ is the heat capacity at constant pressure and can be calculated by $C_p = \sum_{i=1}^{ns} Y_i C_{p,i}$. $p$ satisfies the ideal gas law $p = \rho RT$, where $T$ is the gas temperature, $R$ is the specific gas constant calculated by $R = R_u \sum_{i=1}^{ns} \frac{Y_i}{MW_i}$. $R_u$ is the universal gas constant, and $MW_i$ is the molecular weight of the $i$-th species. $\dot{\omega}_i$ is the net production of $i$-th species. For the chemical reaction with $ns$ chemical components and $nq$ radical reaction equations,

$$\dot{\omega}_i = MW_i \sum_{k=1}^{nq} (v''_{ik} - v'_{ik}) \left[\sum_{i=1}^{ns} (\alpha_{ik} c_{\chi i})\right] \cdot \left[k_{fk} \prod_{i=1}^{ns} (c_{\chi i})^{v'_{ik}} - k_{bk} \prod_{i=1}^{ns} (c_{\chi i})^{v''_{ik}}\right], \quad (5)$$

where $v'_{ik}$ and $v''_{ik}$ are the stoichiometries of the $i$-th species before and after the reaction in the $k$-th reaction, respectively. $\alpha_{ik}$ is the three-body effect coefficient of $i$-th species in $k$-th reaction, $c_{\chi i}$ is the molar concentration of $i$-th species. $k_{fk}$ and $k_{bk}$ are forward and backward reaction rate constants, respectively. Specifically, a skeletal mechanism with 44 species and 112 reactions is used for the simulation of n-heptane detonation [34], which has been demonstrated to have good performance and is widely used in n-heptane/air detonation problems [35, 36]. $S_m$, $\boldsymbol{S}_F$, $S_e$, and $S_{s,i}$ are the source terms from the liquid phase which will be explained in the following text.

The Lagrangian particle tracking (LPT) method is utilized to track the motion of liquid droplets and the changes of droplet masses and temperatures as

$$\frac{dm_p}{dt} = \dot{m}_p, \tag{6}$$

$$\frac{d\boldsymbol{u}_p}{dt} = \frac{\boldsymbol{F}_p}{m_p}, \tag{7}$$

$$c_p \frac{dT_p}{dt} = \frac{\dot{Q}_c + \dot{Q}_l}{m_p}. \tag{8}$$

where $m_p$, $\boldsymbol{u}_p$, and $T_p$ represent the mass, velocity, and temperature of each droplet, respectively. $c_p$ is the constant-pressure heat capacity of the liquid. The droplet is assumed to be a sphere with mass $m_p = \frac{1}{6}\pi\rho_p D_p^3$, where $\rho_p$ and $D_p$ are the density and diameter of the droplet.

The right-hand side terms of Equ. (6) – (8) are calculated by using several sub-models. For droplet mass transfer, the evaporation model presented by Abramzon and Sirignano [37] is used, and the droplet evaporation rate is expressed as

$$\dot{m}_p = -\pi D_p Sh \overline{D}_f \rho_s \, ln(1 + B_M), \tag{9}$$

where $\rho_s$ denotes the fuel vapor density at the surface of the droplet, and $\overline{D}_f$ is the average binary diffusion coefficient of the gas mixture in the films. $Sh$ is Sherwood number modelled by $Sh = 2.0 + 0.6Re_p^{\frac{1}{2}}Sc^{\frac{1}{3}}$, where $Re_p =$ $\frac{\rho_p D_p |\boldsymbol{u}_p - \boldsymbol{u}|}{\mu}$ is the Reynolds number of the droplet, and $Sc = \frac{\nu}{D_f}$ is the Schmidt number. $\nu = \frac{\mu}{\rho}$ is the kinematic viscosity coefficient. The mass transfer number $B_M$ is given by $B_M = \frac{Y_s - Y_g}{1 - Y_s}$, where $Y_s$ and $Y_g$ are the mass fractions of fuel vapor at the droplets' surface and ambient gas, respectively. The value of $Y_g$ is directly obtained from the gas phase conditions, while $Y_s$ is given by $Y_s = \frac{MW_p X_s}{\sum_i X_i MW_i}$, where $MW_p$ is the molecular weight of the fuel vapor, and $X_s$ represents the mole fraction of the fuel vapor on the droplet surface. $X_s$ can be calculated by the Raoult's law as $X_s = X_m \frac{p_{sat}}{p}$, where $X_m$ represents the mole fraction of fuel in the droplet, and $p_{sat}$ corresponds to the saturated pressure of the liquid fuel. The value of $p_{sat}$ is determined using a polynomial function of $T_p$ [38]. Additionally, $\rho_s$ is determined by the Clasius-Clapeyron formula as $\rho_s = p_s MW_m / RT_s$, where $p_s$ is calculated in a similar manner as $p_{sat}$, and $T_s = (T + 2T_p)/3$ represents the temperature on the droplet surface. The droplet force is calculated using sphere drag model as

$$\boldsymbol{F}_p = -\frac{18\mu}{\rho_p D_p^2} \frac{C_d Re_p}{24} m_p(\boldsymbol{u}_p - \boldsymbol{u}), \tag{10}$$

in which

$$C_d = \begin{cases} 24\left(1 + \frac{1}{6} Re_p^{\frac{2}{3}}\right), Re_p \leq 1000 \\ 0.424, \qquad Re_p > 1000 \end{cases}. \tag{11}$$

The droplet convective heat transfer rate is calculated by

$$\dot{Q}_c = h_c A_p (T - T_p), \tag{12}$$

where $A_p = \pi D_p^2$ is the surface area of the droplet, and $h_c = \frac{Nu \cdot k}{D_p}$ is the convective heat transfer coefficient. The Nusselt number $Nu$ is calculated using Ranz-Marshall model [39] as $Nu = 2.0 + 0.6Re_p^{\frac{1}{2}}Pr^{\frac{1}{3}}$, where $Pr = \frac{c_p\mu}{k}$ is the Prandtl number. The droplet evaporation heat transfer rate is expressed as

$$\dot{Q}_l = -\dot{m}_p h_l(T_p), \tag{13}$$

where $h_l(T_p)$ is the latent heat of droplet at $T_p$. Within the above models, the gas-liquid two-way coupling terms can be calculated using PSI-CELL method [40]. The source terms in each calculation cell with volume $V_c$ and $N_p$ particles are expressed as follows,

$$S_m = -\frac{1}{V_c}\sum_1^{N_p} \dot{m}_p\,, \tag{14}$$

$$\boldsymbol{S}_F = -\frac{1}{V_c}\sum_1^{N_p} \boldsymbol{F}_p\,, \tag{15}$$

$$S_e = -\frac{1}{V_c}\sum_1^{N_p} \left(\dot{Q}_c + \dot{Q}_l\right), \tag{16}$$

$$S_{s,i} = \begin{cases} -\frac{1}{V_c}\sum_1^{N_p} \dot{m}_p\,, & for\ condensed\ species \\ 0, & otherwise \end{cases}. \tag{17}$$

In this study, two condensed species are present such as $C_7H_{16}$ and $H_2O$.

## 2.2. *Droplet breakup models*

This study employs four breakup models to examine the impact of these models on the formation of ODWs and the combustion modes in the ODWE combustor: the Taylor analogy breakup (TAB) model [41], the Pilch-Erdman model [42], the ReitzKH-RT model [43], and the dynamic breakup model recently proposed by the authors [44]. Each model encapsulates distinct physical processes, resulting in varied droplet distributions and breakup times.

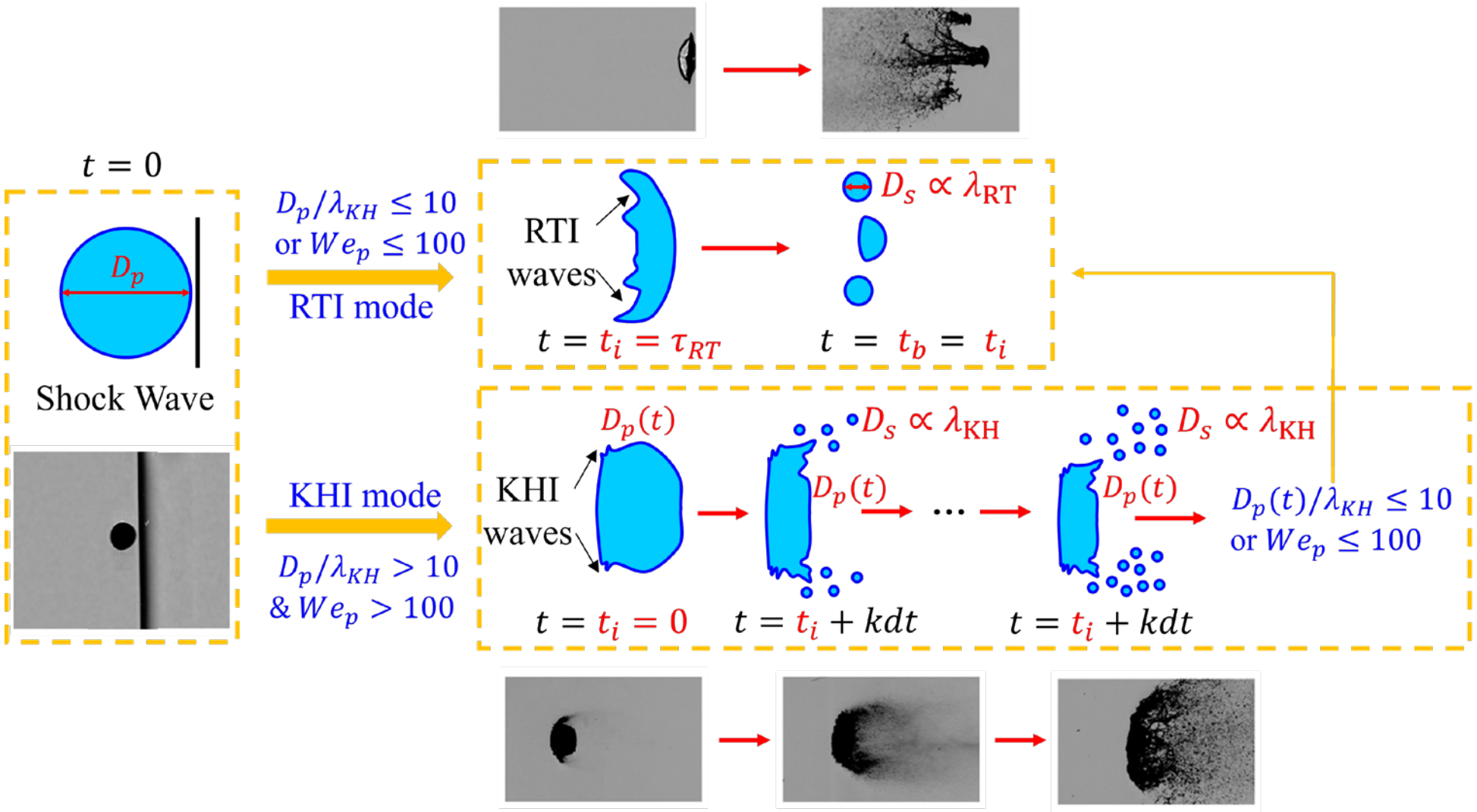

Fig. 1 The underlying physical process of the dynamic model based on experiments [45].

Three key parameters for each model are depicted as follows: the initial breakup time $t_i$, the breakup completion time $t_b$, and the child droplet size $D_s$. The focus here is on the physical processes of the different models, and specific model parameters and equations can be found in [41-44]. The TAB model represents models based on the Taylor analogy, comparing the droplet breakup process to a spring oscillator system. It depicts a transient breakup process where the parent droplet suddenly breaks into smaller droplets upon reaching a certain deformation level. The Pilch-Erdman model represents models based on breakup regions observed in experiments. It describes a chain breakup process, where the diameter of droplet particles in a droplet parcel gradually decreases while their number increases as breakup occurs. This process is governed by an ordinary differential equation, $\frac{dD_p}{dt} = -\frac{D_p - D_s}{\tau}$, where $\tau$ is the breakup time, dependent on the breakup regions. The ReitzKH-RT model elucidates instabilities observed experimentally, depicting the competition between the Kelvin–Helmholtz instability (KHI) and the Rayleigh–Taylor instability (RTI). The KH breakup mode occurs only when the RT breakup mode is absent. The RT breakup mode describes a transient breakup process, whereas the KH breakup mode describes a stepwise breakup process in which child droplets peel off once the parent droplet mass decreases to 97%. This process is also governed by an ordinary differential equation, $\frac{dD_p}{dt} = -\frac{D_p - D_s}{\tau_{KH}}$, in which $\tau_{KH}$ denotes the KH breakup time. However, $\tau_{KH}$ depends on an adjustable parameter, ranging from 1.71 to 100, depending on the specific problem. To eliminate this parameter, the authors proposed a dynamic model that physically models the KHI breakup process, as illustrated in Fig. 1. In this process, the child droplets gradually peel off from the parent droplets. Based on conservation laws, two ordinary differential equations related to the parent droplet diameter $D_p$ and the number of child droplets $n$ are obtained as follows,

$$\begin{cases} \frac{dn}{dt} = f(n, D_p) \\ \frac{dD_p}{dt} = -\frac{\alpha^3 D_p}{3(N + A^3 n)} f(n, D_p) \end{cases}, \tag{18}$$

where

$$f(n(t), D_p(t)) = \frac{\frac{1}{6} N \cdot \rho_l D_p a_p |\boldsymbol{u} - \boldsymbol{u}_p|}{\sigma_p \alpha^2 - \frac{\alpha^3}{3(N + \alpha^3 n)} \left( 2\sigma_p n \alpha^2 + 2\sigma_p N + \frac{N}{2} \rho_l u_p^2 D_p \right)}. \tag{19}$$

In Equ. (20), $a_p$ is the acceleration of the droplet. $\alpha$ is a weak function of $D_p$ expressed as

$$\alpha = 9.02 B_0 \frac{(1 + 0.45 Oh^{0.5})(1 + 0.4 T_a^{0.7})}{\left(1 + 0.87 We_p^{1.67}\right)^{0.6}}, \tag{20}$$

where the Ohnesorge number is defined as $Oh = \frac{\sqrt{We_p}}{Re_p}$, the Taylor number is $T_a = \frac{Oh}{\sqrt{We_p}}$, the droplet Weber number is $We_p = \frac{\rho_l u_r^2 D}{2\sigma_p}$, and $B_0$ is the adjustable parameter from the ReitzKH-RT model. The KH breakup time can be derived from the present model as

$$\tau_{KH} = -\int_{D_0}^{D_s} \frac{3(N + A^3 n)}{f(n, D)A^3} \frac{dD_p}{D_p}, \tag{21}$$

in which $D_0$ is the initial diameter of the droplet. The dynamic breakup model can reduce to the ReitzKH-RT model when $We_p$ is relatively small, and it extends the ReitzKH-RT model to account for higher Weber numbers. Furthermore, this model performs well in predicting liquid-fueled detonation parameters, yielding detonation cell sizes similar to those of the ReitzKH-RT model, without requiring the KH breakup time parameter [44].

### 2.3. *Computational specifications*

In this work, a prototypical combustor of a pre-injection ODWE [6] is selected to investigate the formation of ODWs. This ODWE is designed to operate at Mach 9 with a two-stage inlet, featuring a two-dimensional geometry, as shown schematically in Fig. 2 (a). The computational domain encompasses the combustor of this ODWE, with its detailed structure and boundary conditions illustrated in Fig. 2 (b). The combustor measures 300 mm in length and 60 mm in height, with an upper wall angled at 25° relative to the inflow. The upper wall functions as a wedge to induce the OSW and subsequent ODW in the fuel-air mixture inflow. To achieve the successful formation of the ODW, a combustor with an on-wedge strip, as depicted in the green dotted box in Fig. 2 (b), is also designed in accordance with the experiment by Han et al. [33]. Strip diameters of 2 mm and 5 mm are employed for comparison.

For the boundary conditions, the inflow conditions of the combustor are determined by assuming that the ODWE operates at a Mach number of 9 and an altitude of 30 km. The air is compressed by two-stage OSWs in the inlet, with a deflection angle of 12.5° for each stage, and is premixed with n-heptane fuel at stoichiometric ratios before entering the combustor. Consequently according to the theoretical solutions of oblique shock waves, the inflow pressure is $1.963\times10^5$ Pa, the temperature is 814.4 K, and the velocity is 2466 m/s at the entrance of the combustor. The n-heptane droplets are injected randomly from the inflow boundary with the same velocity as the air, and are monodispersed with a diameter of 10 μm and a temperature of 300 K. This work primarily addresses the formation of ODWs and combustion modes within the ODWE combustor, with the influence of the boundary layer on ODW stability to be explored in future research. Therefore, on the upper and lower walls of the combustor, a slip-reflecting boundary condition is employed due to the very thin boundary layer resulting from the high Reynolds number, which is consistent with that adopted in previous studies [32, 46]. For the far-field and outflow boundaries, the zero-gradient Neumann boundary conditions are adopted for pressure, temperature, and velocity.

The gas-phase and liquid-phase equations are solved using an open-source two-phase supersonic reactive flow solver developed by the authors [44] based on OpenFOAM V7 [47]. This solver incorporates chemical reactions, multi-component transport, and a liquid-phase Lagrangian solver into *rhocentralfoam*, and has been validated in one- and two-dimensional liquid-fueled normal detonation problems. In this solver, the finite volume method (FVM) is employed to solve the governing equations of the gas phase. The first-order Euler scheme is used for the temporal discretization, and the "Gauss-Limited linear" scheme is used for spatial derivatives. The CFL number is set as 0.4 for the gas flow. For the convective term, the KNP scheme is utilized to capture the shock wave [48].

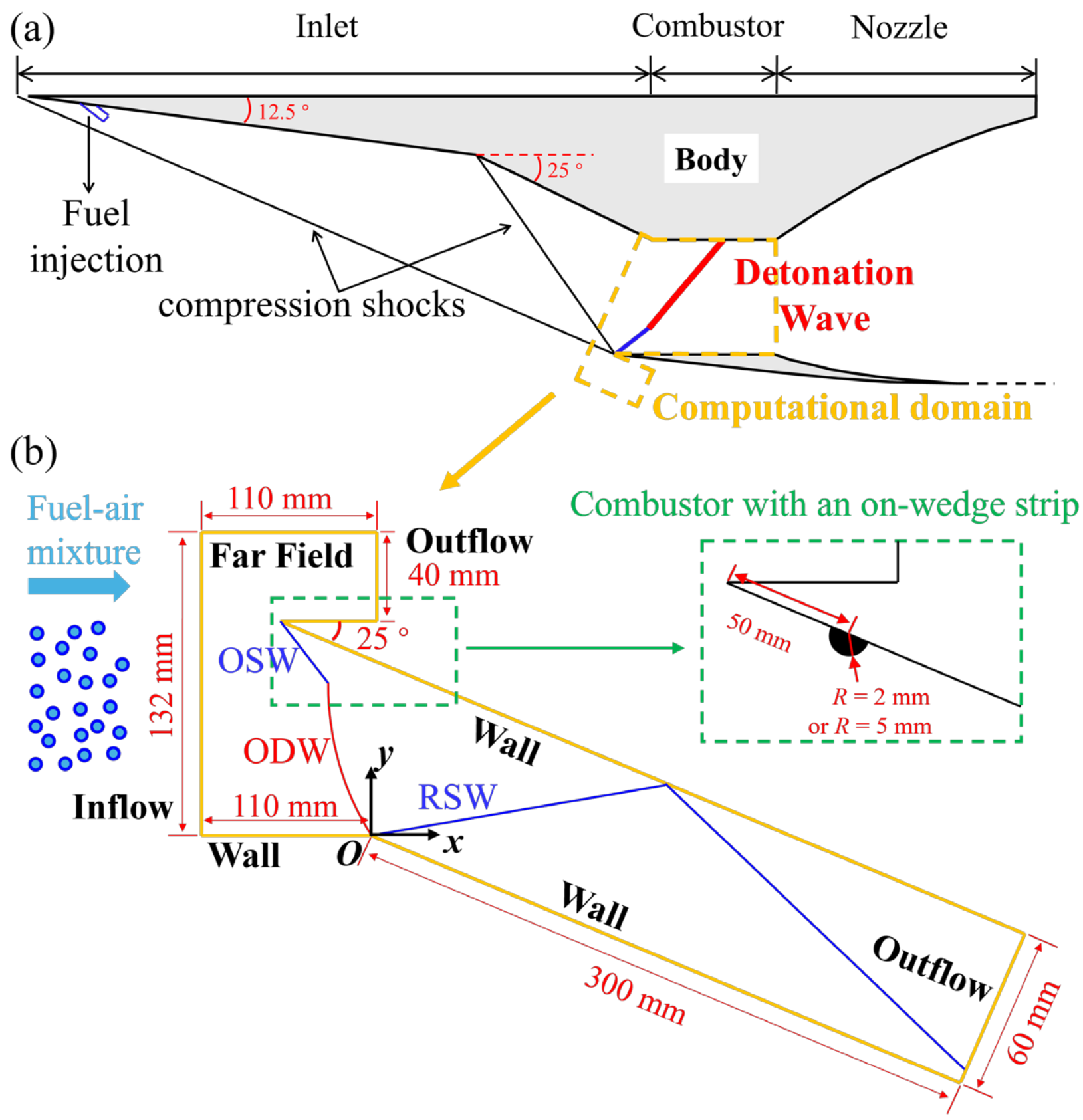

Fig. 2 (a) Schematic of a prototypical pre-injection oblique detonation wave engine (ODWE). (b) The schematic, dimensions, and boundary conditions of the computational domain in this ODWE. The green dotted line represents the different configuration of the combustor with the introduction of the on-wedge trip. Note that the length in the figure does not represent the actual domain.

### 2.4. *Mesh-independence verification*

To demonstrate the mesh independence of the present two-phase ODW simulations, a series of simulations were conducted with varying mesh sizes for the ODW problem in the combustor. Three mesh sizes employed were 0.2 mm, 0.3 mm, and 0.4 mm, corresponding to the total number of meshes being $6.94\times10^5$, $3.47\times10^5$, and $1.96\times10^5$, respectively. The simulations used an inflow of n-heptane vapor/air mixture, with streamlines extracted near the upper wall of the combustor, passing through the same reference point when the flow field stabilized at $t = 1$ ms. Figure 3 displays the flow parameters along these streamlines for various mesh sizes. The curves indicate that the temperature, pressure, and $C_7H_{16}$ mass fraction from these meshes are almost the same. Consequently, a mesh size of 0.3 mm as a balance of computational accuracy and load is selected for subsequent parametric studies. This mesh resolution is also consistent with other research on wedge-induced n-heptane liquid-fueled problems [32, 49].

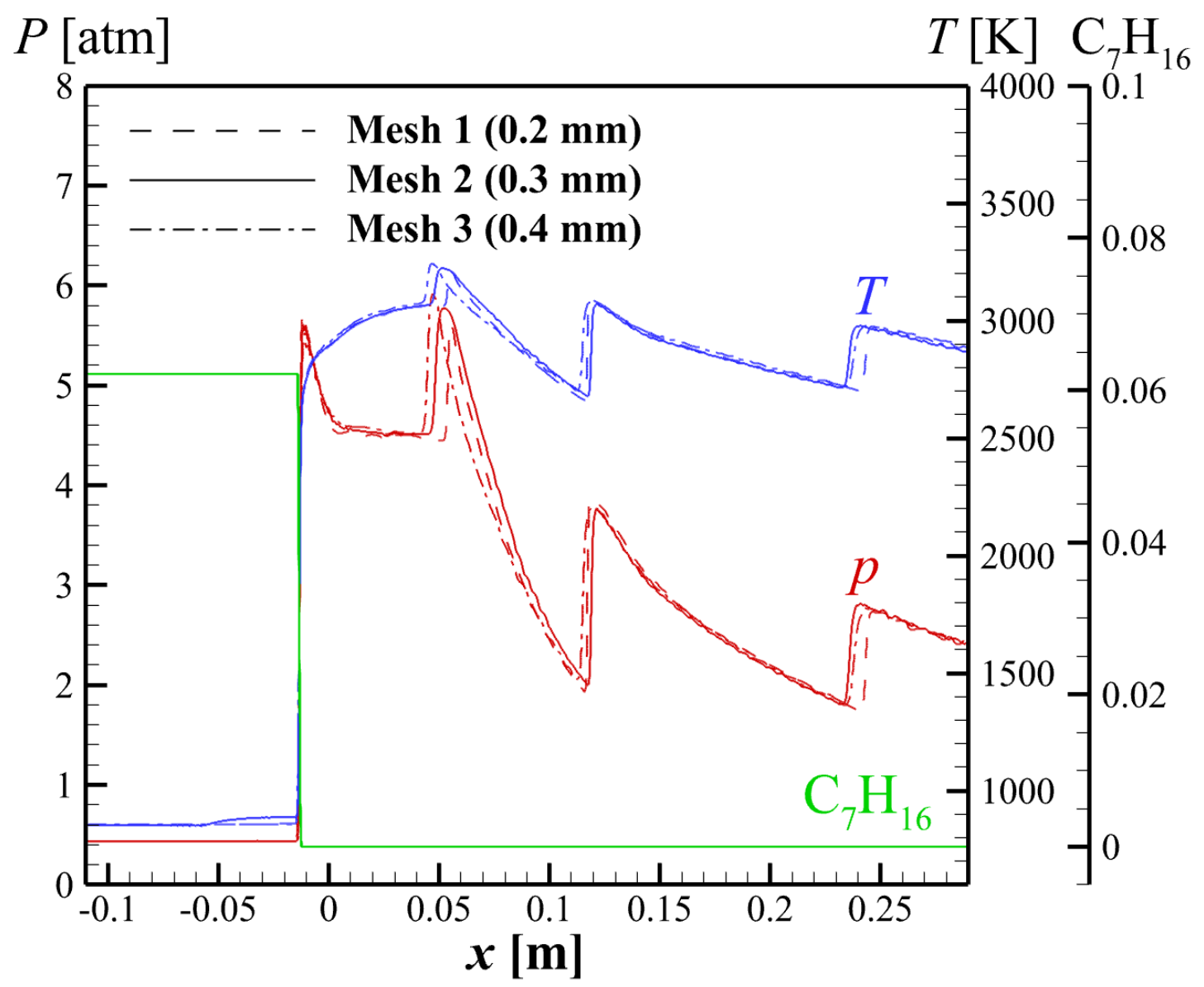

Fig. 3 The flow parameters along streamlines that pass through the same point for cases with n-heptane vapor/air inflow and various mesh sizes. The total number of meshes are $6.94\times10^5$, $3.47\times10^5$, and $1.96\times10^5$, respectively.

## 3. Results and Discussion

### *3.1. Wedge-induced ODW: Numerical validation*

The solver employed in this work has been validated against experimental data for one- and two-dimensional liquid-fueled normal detonation wave problems in the previous paper [44]. For the liquid-fueled ODW over wedges, experiments have rarely been reported. To further validate the numerical methods for two-dimensional ODW problems discussed in this paper, we adopted the three typical cases numerically studied by Teng et al. [32, 50], namely Case 1: gaseous hydrogen ODW, Case 2: gaseous n-heptane ODW, and Case 3: ODW in a mixture of 70% n-heptane vapor and 30% n-heptane droplets over wedges. For Case 1, the inflow pressure is $1.963\times10^5$ Pa, and the inflow temperature is 814.4 K; for Case 2 and 3, the inflow pressure is $2.855\times10^4$ Pa, and the inflow temperature is 697.0 K. The fuel-air stoichiometric ratio is 1 for all three cases. The schematic of the two-dimensional computational domain and boundary conditions for the validation cases are shown in Fig. 4, where the fuel-air mixture passes through a wedge, forming an OSW that transitions into an ODW. The left and upper boundaries are set as inflow with stoichiometric fuel-air mixtures; the right boundary is set as supersonic outflow with a zero-gradient condition for all flow variables; and the lower boundary is set as a slip-reflecting wall. For Case 1 with hydrogen, the present solver employs a skeletal mechanism with 9 species and 21 reactions [51], while Teng et al. [50] adopt a mechanism with 13 species and 27 reactions [52]. For Cases 2 and 3 with n-heptane, both the present solver and Teng et al. [32] utilize a skeletal mechanism comprising 44 species and 112 reactions.

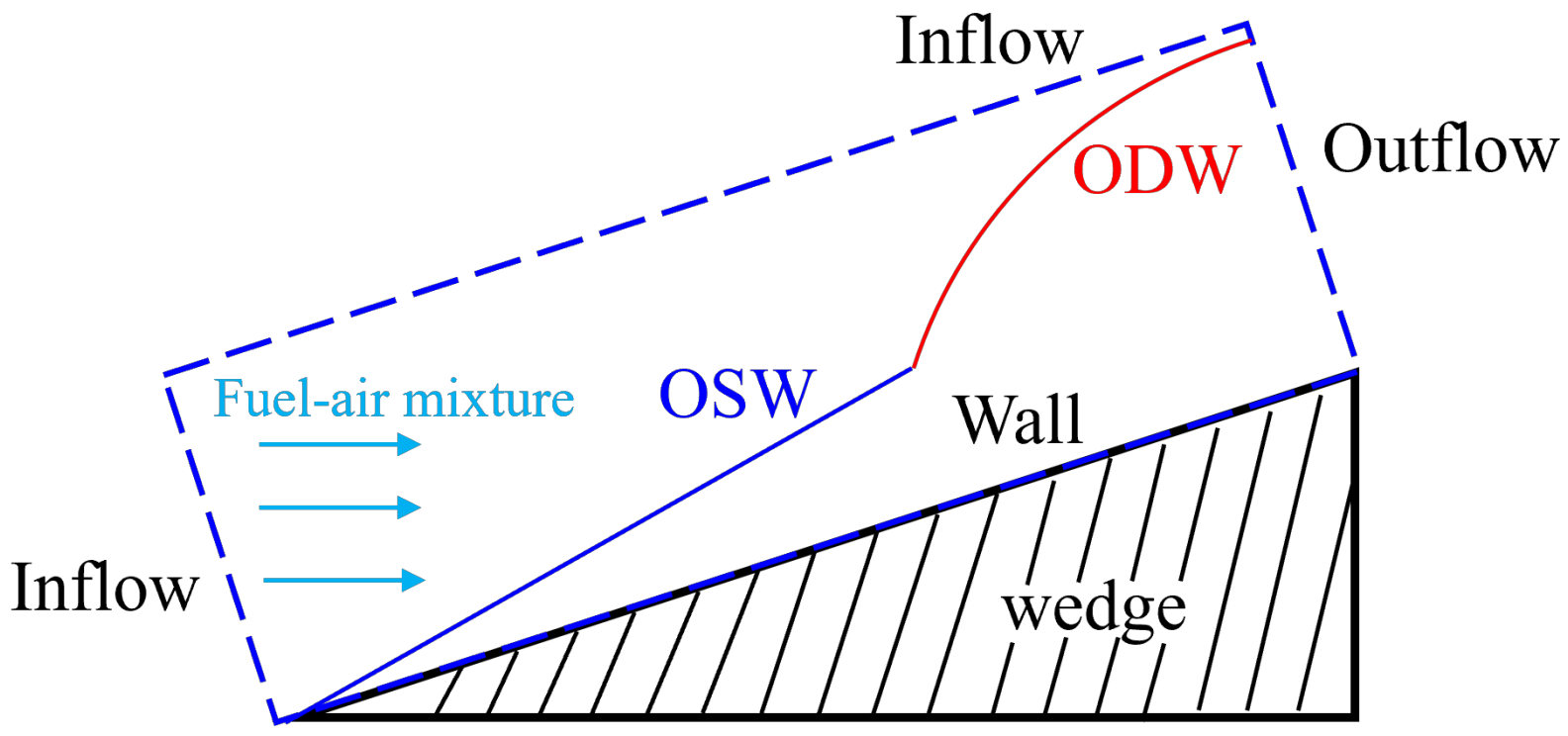

Fig. 4 Schematic of the computational domain and boundary conditions for the two-dimensional wedge-induced oblique detonation wave (ODW) in a fuel-air mixture. The OSW in the figure represents the oblique shock wave.

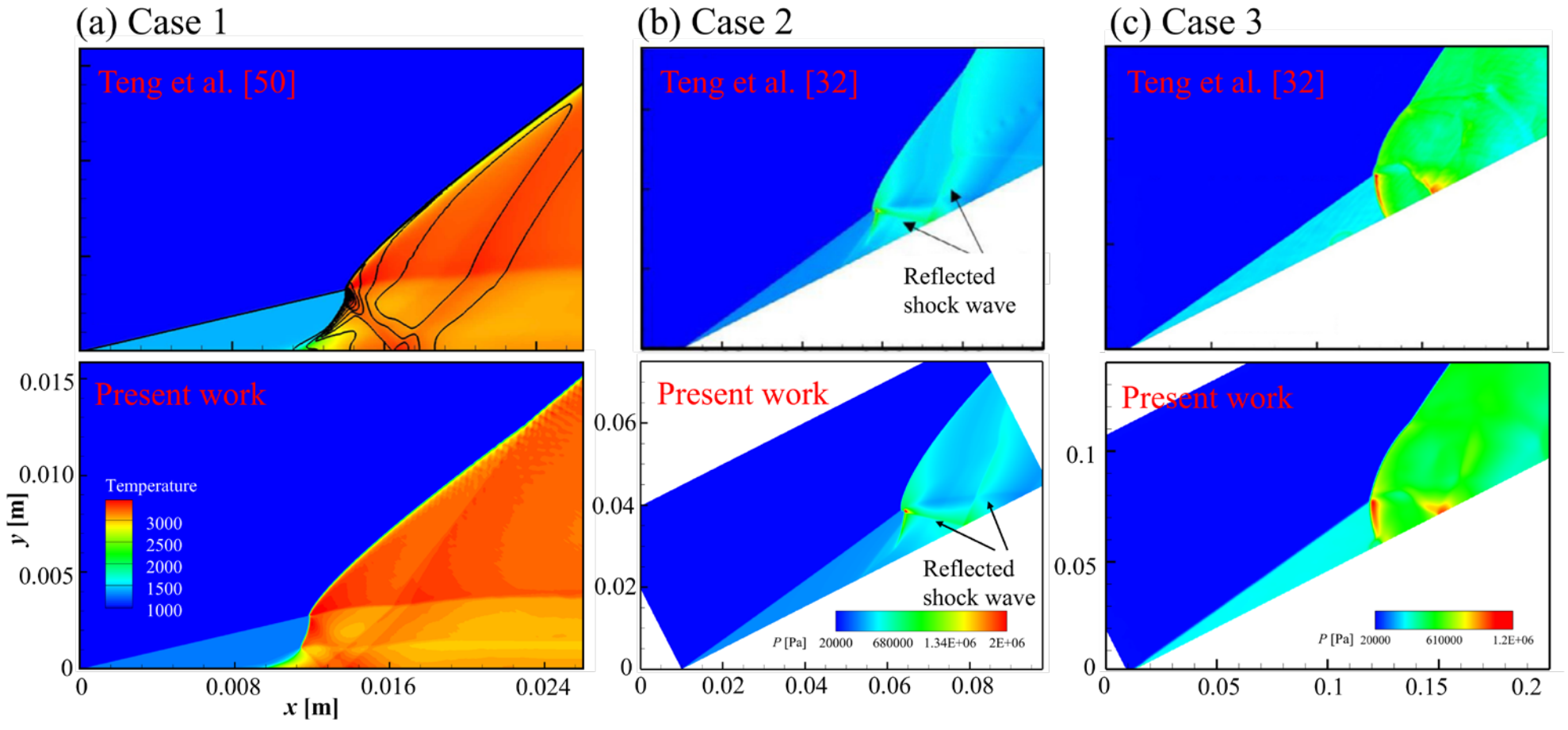

Fig. 5 Comparison of temperature and pressure contours between Teng et al.'s work [32, 50] and the present solver. (a) Temperature contour with hydrogen gas/air inflow. (b) Pressure contour with n-heptane vapor/air inflow. (c) Pressure contour with n-heptane vapor/droplets/air mixture inflow.

Figure 5 (a) illustrates the temperature contours at steady state for Case 1. The present solver predicts a similar ODW morphology as that of Teng et al. [50], classifying this structure as the Type III wave system, an abrupt transition mode. Three sections are discernible in the figure: compression waves (CWs) near the wall, a secondary oblique detonation wave (SODW) in the middle, and a main ODW at the top. Together, these three sections constitute the reactive front. For quantitative comparison, three detonation parameters of this case are shown in Table 1. The initiation length is defined as the distance from the front tip to the end of the induction zone, where the temperature increases to 110% of the average post-shock temperature; the OSW angle and ODW angle are the angles between the wave and the wall. Table 1 demonstrates that the present solver predicts detonation parameters similar to those of Teng et al. [50], with a discrepancy of 8.93% for the initiation length and discrepancies less than 4.00 % for the OSW and ODW angles. The difference in initiation length may result from the different skeletal mechanisms employed. Despite this, the differences in wave angles are quite small.

Figure 5 (b) and (c) represent the pressure contours for Cases 2 and 3. The only difference between these two cases is the gas-liquid ratios. For Case 2, the present solver gives the same ODW morphology as that of Teng et al. [32], and this structure can also be classified as the Type III wave system. The multi-wave point exhibits a relatively high pressure, from which a reflected shock wave (RSW) forms and interacts with the wall. The detonation parameters in Table 1 show that the errors between these two simulations are less than 3%, indicating that, using the same skeletal mechanism, the present solver can accurately predict the oblique detonation parameters as Teng et al. [32].

For Case 3, with the introduction of the n-heptane droplets, the OSW reaches a steady state, but the subsequent ODW remains unstable, fluctuating within a certain range. Therefore, the pressure contours for the same state in one floating cycle for the results of this solver and Teng et al. [32] are extracted and shown in Fig. 5 (c). Both simulations predict the abrupt transition mode and the same high-pressure area. The detonation parameters are presented in Table 1. Due to the unstable flow field, the initiation length fluctuates over time, so average initiation lengths are calculated, with the error being less than 3%. Additionally, the wave angles in Fig. 5 (c) are measured and compared in Table 1, with errors of less than 4%. The above three comparisons demonstrate the accuracy of the present solver in calculating gaseous and liquid-fueled

ODW problems.

| Case | Types | Initiation length | Initiation length error | OSW angle | OSW angle error | ODW angle | ODW angle error |
|---|---|---|---|---|---|---|---|
| 1 | Teng et al. [50] | 0.112 m | - | 13.58° | - | 36.96° | - |
| | Present work | 0.102 m | 8.93% | 13.37° | 1.55% | 38.44° | 4.00% |
| 2 | Teng et al. [32] | 0.042 m | - | 9.17° | - | 25.32° | - |
| | Present work | 0.041 m | 2.38% | 9.02° | 1.64% | 25.06° | 1.03% |
| 3 | Teng et al. [32] | 0.128 m | - | 8.65° | - | 30.32° | - |
| | Present work | 0.131 m | 2.29% | 8.43° | 2.54% | 29.20° | 3.69% |

Table 1 The initiation length, OSW angle, and ODW angle for the wedge-induced ODWs for different validation cases. The results predicted by the present work and Teng et al.'s studies [32, 50] are shown together for comparison.

## 3.2. *Formation of ODWs in engine combustor*

In this section, the formation of ODWs in the combustor and the influence of various factors on their formation are investigated using the control variable method. First, a benchmark case is analyzed, wherein the inflow contains only n-heptane droplets and air, with no on-wedge strips, employing the Pilch-Erdman droplet breakup model. Subsequently, the effects of different droplet breakup models, the presence of n-heptane vapor, and the implementation of on-wedge strips are examined separately.

### *3.2.1 Phenomenology and wave structures of benchmark case*

Figure 6 presents the numerical schlieren graph within the combustor for the benchmark case at different time points, depicting the establishment of the flow field. When the fuel/air mixture reaches the combustor and encounters the upper wall, an OSW immediately forms over the upper wall, but it does not transition into an ODW. Conversely, as the mixture passes along the lower wall, a low-velocity zone forms due to the expansion of the flow channel, resulting in the gradual formation of a separation shock wave (SSW). Additionally, the flow field and wave structures in the combustor have reached a relatively stable state at $t = 1$ ms.

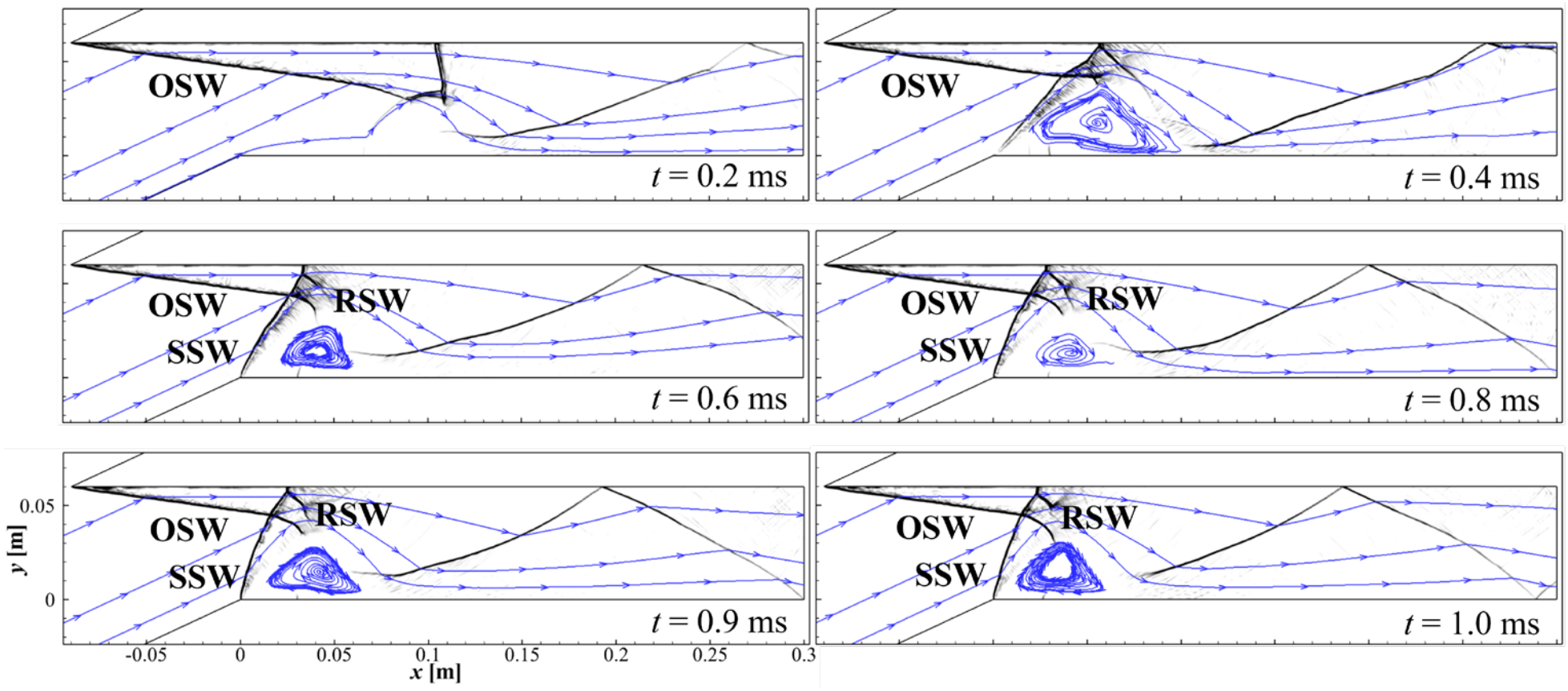

Fig. 6 The establishment of flow filed in the combustor for the benchmark case. For clarity, the flow field has been rotated counterclockwise by 25°.

More detailed flow fields at $t = 1$ ms are shown in Fig. 7 (a) – (c). Figure 7 (a) presents the numerical schlieren graph of the combustor, where the red line represents the sonic line. This figure illustrates that the OSW and SSW intersect, creating two reflected shock waves (RSWs), one of which undergoes a Mach reflection at the wall, forming a Mach stem (MS). Figure 7 (b) and (c) illustrate the n-heptane vapor mass fraction contour and temperature contour within the combustor. The droplet distribution is also depicted in these figures, with droplets proportionally enlarged for clarity. The figures indicate that n-heptane droplets evaporate slowly after entering the combustor, so there is only a small increase in the n-heptane vapor mass fraction. Post-OSW, droplets rapidly break up and evaporate into n-heptane vapor. However, the n-heptane vapor does not participate in gaseous combustion until it encounters the MS and the RSW, which elevate the temperature of the n-heptane vapor/air mixture. Droplets passing through the SSW undergo breakup, evaporation, and combustion in a thin region, resulting in nearly zero n-heptane vapor mass fraction beyond the SSW.

These results demonstrate that, under the specified inflow conditions, the combustor cannot operate as intended: the ODW is initiated over the upper wall and reflected at the lower wall. In this configuration, the wedge-induced OSW fails to induce the ODW. Figure 7 (d) presents the experimental schlieren graph by Han et al. [33]. In the experiment, two parallel strut injectors were used to inject RP3 kerosene before the inlet. The kerosene spray undergoes one-stage compression from the OSW, mixes with the air, and then enters the combustor. The inflow Mach number at the inlet fluctuates around 6.5. Although we employed different fuel types and inlet conditions in the numerical simulations due to the complexity and uncertainty of experimental conditions, which hampers a quantitative comparison between experimental and numerical results, comparing the underlying flow structures remains insightful. The ODW combustion mode is absent in both the present numerical solutions and the reported experimental results [33]. However, numerical results predict combustion after the SSW, while experiments show no high-temperature area post-SSW. This discrepancy may arise from differences in fuel injection methods: in the experiments, fuel is injected horizontally into the mainstream, resulting in less fuel near the lower wall and a diminished likelihood of reaction. Conversely, simulations assume a random distribution of fuel, which may allow reactions to occur post-SSW. Despite these differences, the numerical simulations effectively capture the fundamental physical processes and flow structures observed in the experiments.

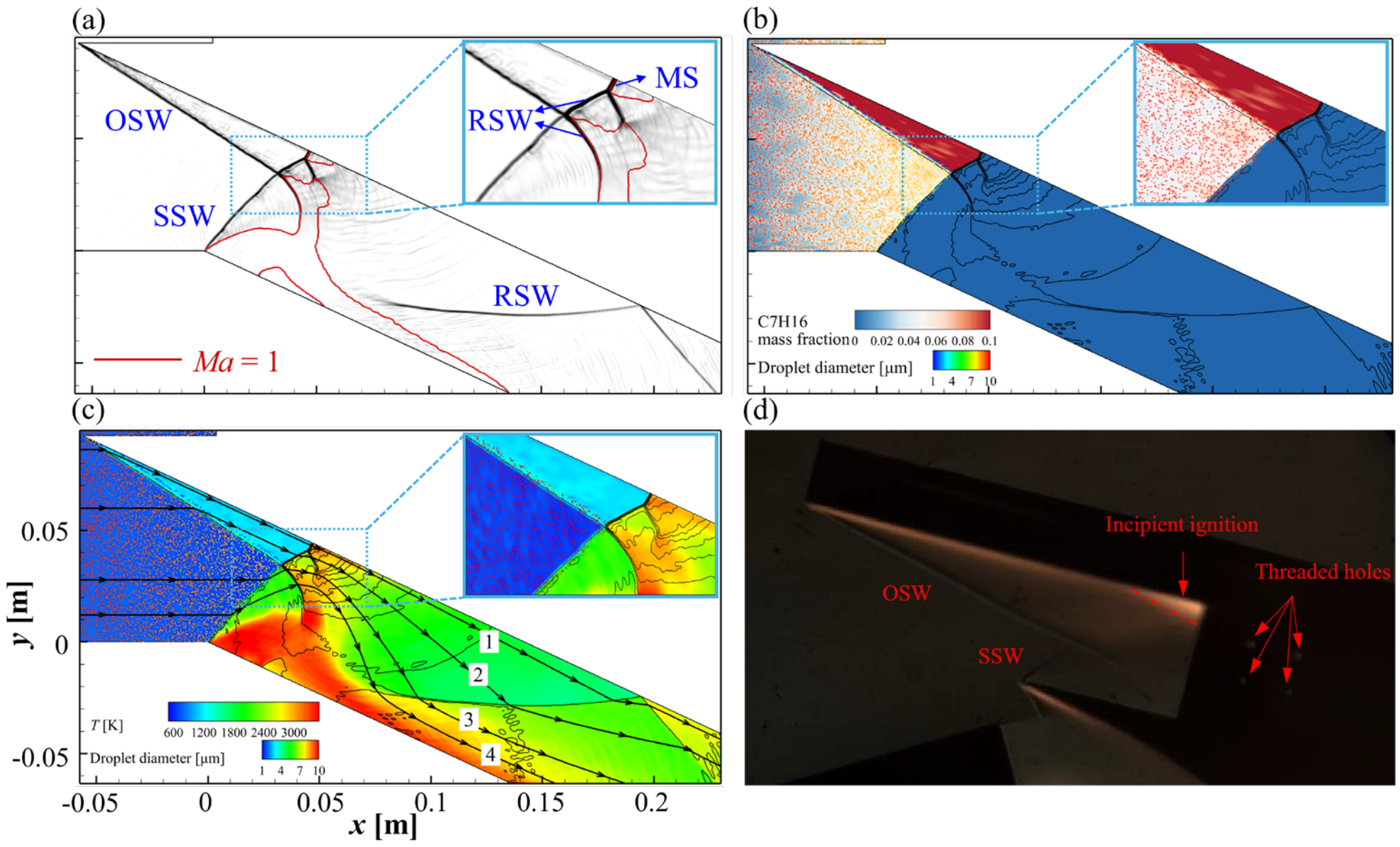

Fig. 7 (a) Numerical schlieren graph, (b) $C_7H_{16}$ mass fraction contour with droplet distribution, and (c) temperature contour with droplet distribution in the combustor at $t = 1\,ms$ for the benchmark case. The droplet sizes are magnified for clarity. (d) Experimental schlieren graph by Han et al. [33]. SSW: separation shock wave; RSW: reflected shock wave; MS: Mach stem.

To further understand the detailed flow field in the combustor, the pressure, temperature, and n-heptane vapor mass fraction along four representative streamlines, as shown in Fig. 7 (c), are plotted in Fig. 8. Figure 8 (a) and (b) depict the parameters of streamlines passing through the OSW induced by the upper wall. The distinction between these two streamlines is that Streamline 1 passes through the MS, while Streamline 2 passes through the RSW. The curves illustrate that, before the OSW, n-heptane droplets gradually evaporate into vapor, resulting in a gradual temperature decrease. Post-OSW, the n-heptane vapor mass fraction rapidly increases due to droplet breakup after the shock wave and accelerated evaporation prompted by the elevated temperature. The temperature also increases abruptly after the OSW, then slightly diminishes due to heat absorption from evaporation. At this temperature, the reaction does not occur when the mixture flows parallel to the wall, with flow parameters fluctuating marginally until the MS or RSW, indicating that the OSW cannot transition into ODW in the confined space. Subsequently, with the substantial rise in temperature and pressure caused by the MS or RSW, the n-heptane vapor reacts swiftly within a narrow region in the streamwise direction.

Figure 8 (c) and (d) represent the parameters of streamlines passing through the SSW induced by the low-velocity zone. Compared to Streamlines 1 and 2, Streamlines 3 and 4 travel a longer distance before the shock wave, resulting in a longer evaporation period and a larger temperature reduction from approximately 800 K to 600 K. Post SSW, the temperature abruptly increases to above 2000 K, and the n-heptane vapor mass fraction precipitously diminishes to zero, signifying that the n-heptane droplets preceding the SSW undergo instantaneous breakup, evaporation, and gaseous combustion near the SSW. However, the high-temperature area is decoupled from the SSW, signifying shock-induced combustion in this region.

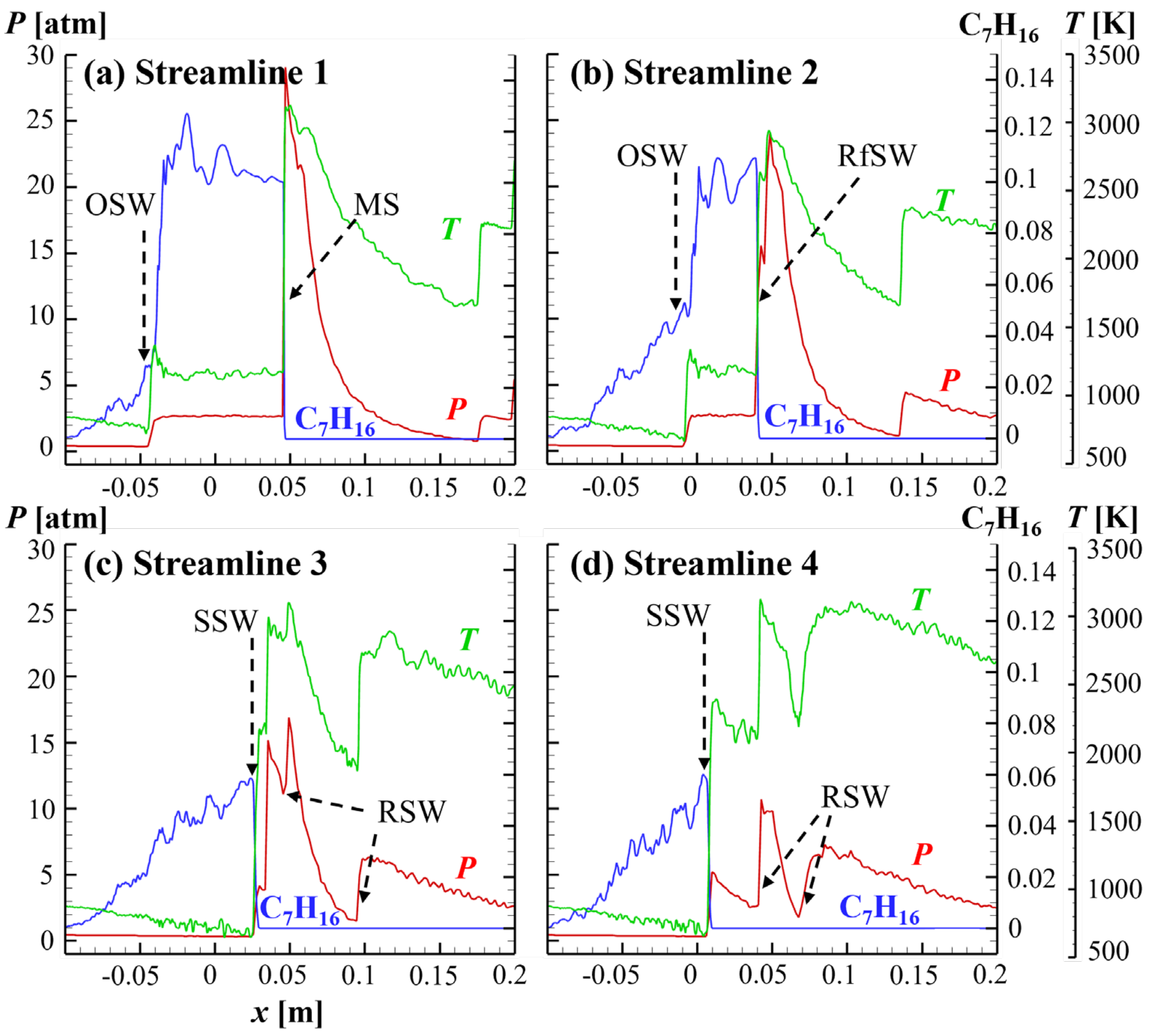

Fig. 8 Flow parameters along four representative streamlines at $t = 1\,ms$ in the combustor for the benchmark case.

### *3.2.2 Effects of droplet breakup models*

In this section, the benchmark case is computationally reproduced by using three different breakup models, namely the TAB model, the ReitzKH-RT model, and the dynamic model, to investigate their influence on the flow field structures. Figure 9 illustrates the flow structure for different breakup models at 1 ms. The left figures display the numerical schlieren images, with the sonic lines marked in red, while the right figures show the temperature contours and droplet distribution.

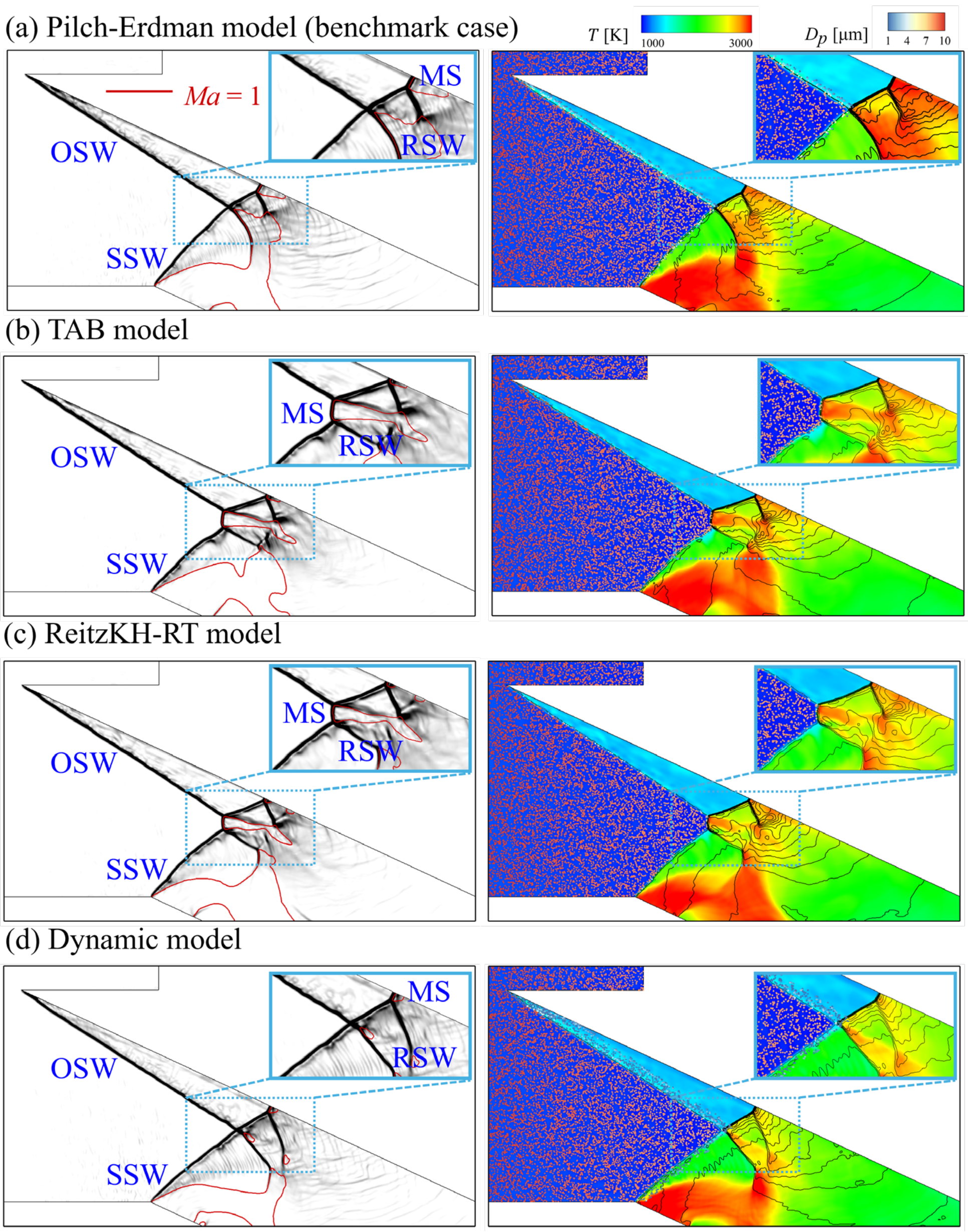

Fig. 9 Flow field in the combustor with pure n-heptane droplets/air mixture inflow using different droplet breakup models. (a) Pilch-Erdman model (benchmark case), (b) The TAB model, (c) ReitzKH-RT model, and (d) dynamic model. The left panels display numerical schlieren graphs, while the right panels present temperature contours with droplet distributions. The droplet sizes are magnified for clarity.

In Fig. 9 (b), the TAB model shows slight deviations from the benchmark case depicted in Fig. 9 (a). As in the benchmark, the OSW fails to transition into the ODW over the upper wall. However, upon interaction with the SSW, an irregular reflection occurs, forming a MS. In the ReitzKH-RT model, shown in Fig. 9 (c), the flow field closely resembles that of the TAB model, with a MS forming at the interaction point. Additionally, the rapid disappearance of droplets soon after the OSW suggests a quick breakup process. Conversely, the dynamic model in Fig. 9 (d) aligns well with the benchmark case, with droplets traveling a certain distance before completely breaking up and evaporating into n-heptane vapor, and no MS forming at the interaction point.

To elucidate the differences between these models, the angles between the OSW and the SSW were measured. The TAB and ReitzKH-RT models predict angles of 78.7° and 79.2°, respectively, while the benchmark case and the dynamic model yield angles of 75.9° and 73.1°, respectively. The larger angles observed in the TAB and ReitzKH-RT models result in irregular reflections. This phenomenon may be attributed to the rapid breakup of droplets in these models as the parent droplets pass through the OSW, leading to stronger reactions post-SSW and thereby enlarging the angle between the ODW and the SSW.

Despite the minor differences at the intersection point of the two shock waves, the ODW cannot form in the combustor under the given conditions, regardless of the breakup model employed. This indicates that the breakup process is unlikely the primary reason for the unsuccessful formation of the ODW. The dynamic model yields result most similar to that of the Pilch-Erdman model, which is also frequently used in liquid-fueled oblique detonation problems [32, 49]. However, the dynamic model incurs a larger computational cost due to the introduction of new droplet parcels. Therefore, the Pilch-Erdman model will be employed in subsequent simulations.

*3.2.3 Effects of gas-liquid ratios*

The benchmark case shows that, under the given conditions, the ODW fails to form in the n-heptane droplets/air mixture in the combustor. Tian et al.'s numerical simulations on n-heptane-fueled oblique detonation over a wedge have demonstrated that maintaining a total equivalence ratio of 1 and keeping the droplet diameter below 20 μm, an increase in the proportion of n-heptane vapor will shorten the initiation length [49]. Therefore, in this section, we will follow the same approach by introducing n-heptane vapor to the spray to enhance the formation of ODWs. Two additional gas-liquid ratios are adopted: one with pure n-heptane vapor and one with a mixture of 50% n-heptane vapor and 50% droplets.

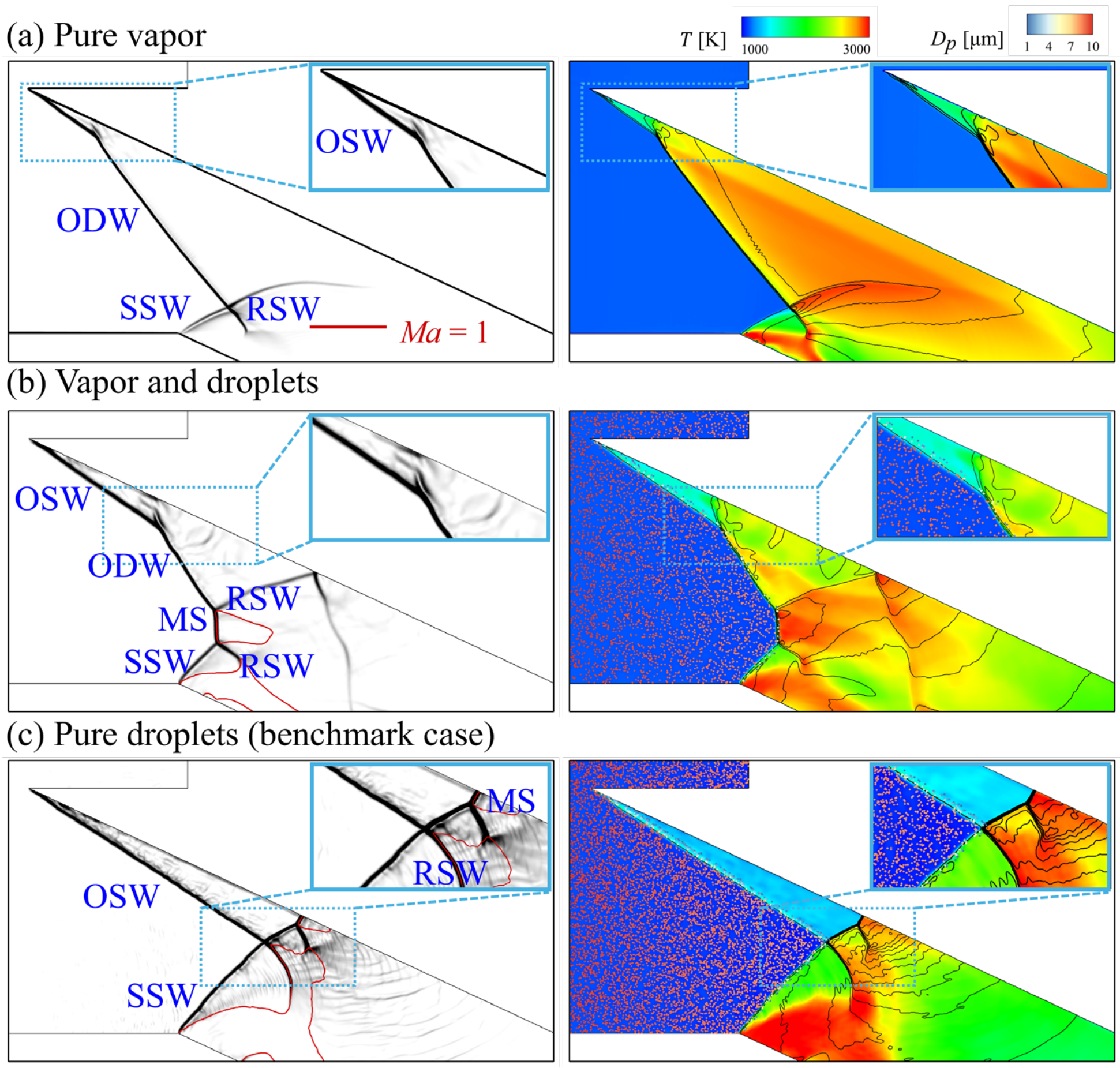

Fig. 10 Flow fields in the combustor under various n-heptane gas-liquid ratios without an on-wedge strip: (a) Pure n-heptane vapor, (b) a mixture of n-heptane vapor and droplets, and (c) pure n-heptane droplets (benchmark case). The left panels display numerical schlieren graphs, while the right panels present temperature contours with droplet distributions. The droplet sizes are magnified for enhanced visualization.

As shown in Fig. 10 (a), for the case with a pure n-heptane vapor/air mixture as the inflow condition, the OSW induced by the upper wall successfully transitions into an ODW, leading to successful combustion and subsequent high post-wave temperatures. Due to the ignition delay time of n-heptane vapor, the transition point is at a certain distance from the OSW's initiation point. Consequently, the ODW may not reflect at the lip of the lower wall of the combustor, forming a small low-velocity zone and resulting in the formation of an SSW. Nevertheless, the ODW combustion occupies most of the combustor, as expected from the combustor design. For the case with the n-heptane vapor/droplets/air mixture as the inflow condition, illustrated in Fig. 10 (b), the ODW also successfully forms on the upper wall. However, the transition length from OSW to ODW is notably longer, and the area of ODW combustion is significantly smaller than in the case of pure n-heptane vapor. Furthermore, when the ODW interacts with the SSW, a Mach reflection occurs, resulting in a MS. This MS is stronger than the ODW, with a higher post-wave temperature, allowing the n-heptane to react near the wave. Therefore, this MS represents another detonation combustion mode. Consequently, ODW and MS combustion modes coexist in this case. For the benchmark case with pure droplets, shown in Fig. 10 (c), as discussed in the previous section, the OSW fails to transition to ODW within the confined length. The comparison indicates that the ODW can successfully form with the presence of n-heptane vapor.

To analyze the formation mechanism of ODW with the introduction of n-heptane vapor, the flow parameters, including temperature and key chemical reaction species along streamlines near the upper wall of these three cases, are extracted and plotted in Fig. 11. These streamlines pass through the same starting point as Streamline 1 in Fig. 7 (c). They are chosen because the fuel/air mixture will travel the shortest distance before encountering the OSW, thus minimizing the effect of pre-evaporation of the n-heptane droplets. Additionally, the mixture along these streamlines will pass through the OSW before participating in any reactions, ensuring that the transition process is decoupled. Here, chemical reactions and mixture flow occur simultaneously, so changes in chemical reaction species over time are depicted as changes over space. For the skeletal mechanism used in the present work, n-heptane is the reactant, and $H_2O$ is one of the ultimate products. Three other reactions are considered: the chain initiation reaction and two key chain branching reactions. One chain branching reaction is the dissociation of keto-heptyl peroxide (KET),

$$\mathrm{ORO_2H\ (KET) \rightarrow ORO + OH}, \tag{22}$$

which has relatively high activation energy, and the buildup of KET signals low-temperature chemistry. The other is the decomposition of $H_2O_2$,

$$\mathrm{H_2O_2 + M \rightarrow 2OH + M}, \tag{23}$$

in which the decomposition of $H_2O_2$ represents intermediate-temperature chemistry.

For the case with pure n-heptane vapor, as depicted by the red lines, after the OSW, the temperature increases, and the initiation reaction occurs immediately, consuming $C_7H_{16}$ and producing $C_7H_{15}$. No KET is built up in the entire process, indicating the absence of low-temperature chemistry. In contrast, $H_2O_2$ gradually accumulates, and a small amount of $H_2O$ is generated slowly before $x = -0.04$ m. Then, $H_2O_2$ is quickly consumed, meaning that intermediate-temperature chemistry occurs. Simultaneously, $H_2O$ is rapidly produced, and the temperature increases sharply as $C_7H_{16}$ is depleted, indicating that the heat release reactions occur swiftly. Consequently, with rapid chemical reactions, the OSW successfully transitions into the ODW, and this process is dominated by intermediate-temperature chemistry.

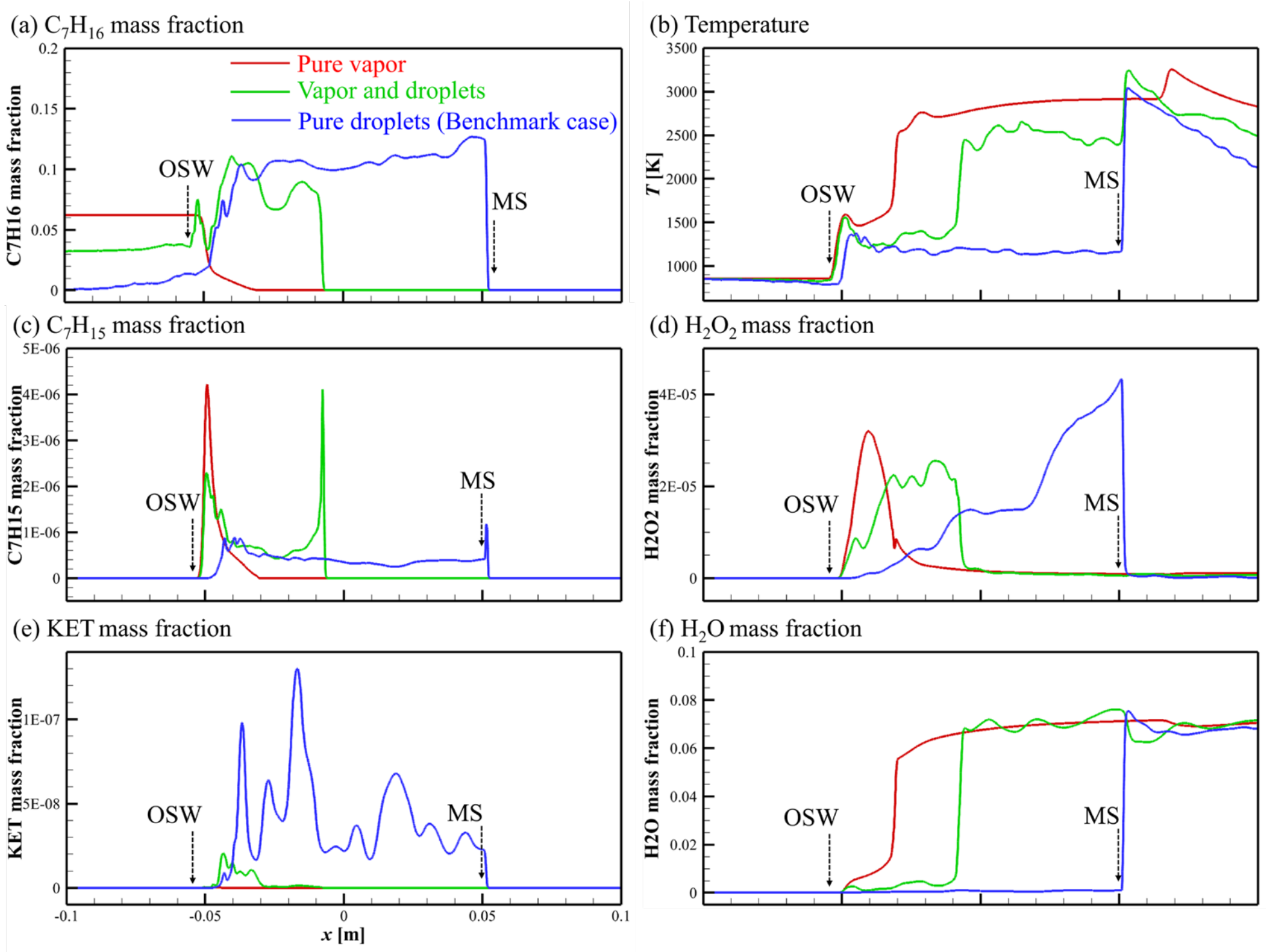

Fig. 11 Flow parameters along streamlines passing through a specific point near the upper wall with various gas-liquid ratios.

The chemical reaction behaves differently in the case of pure n-heptane droplets, depicted by the blue lines. After the OSW, the temperature increases, leading to rapid evaporation of n-heptane droplets and the subsequent increase in $C_7H_{16}$ vapor mass fraction. Due to this evaporation process, the $C_7H_{15}$ mass fraction increases after a certain length post-OSW, indicating that the initiation reaction is slightly delayed. With the initiation of the chemical reaction, KET builds up continuously, and $H_2O_2$ increases steadily, indicating the presence of low-temperature chemistry and the absence of intermediate-temperature chemistry. This situation persists over a long distance until the mixture passes through the MS caused by the RSW. After the MS, the temperature abruptly increases to above 3000 K. At this temperature, $H_2O_2$ sharply decreases, and $H_2O$ is quickly produced, indicating that intermediate-temperature chemistry and heat release reactions occur. Nonetheless, this process is induced by the high temperature after the MS instead of the transition from OSW to ODW.

In the case of a mixture of n-heptane droplets and vapor, shown in green lines, a certain amount of $C_7H_{16}$ vapor exists before the OSW. After the OSW, the temperature increases similarly to the case with pure n-heptane vapor, resulting in an increase in n-heptane vapor mass fraction. Due to the presence of $C_7H_{16}$ vapor, the initiation reaction occurs immediately post-OSW. As the reactions commence, KET gradually builds up, and the $H_2O_2$ mass fraction increases, indicating the presence of low-temperature chemistry and the absence of intermediate-temperature chemistry. However, this situation only persists for a short distance. At $x = -0.3$ m, the KET mass fraction gradually decreases to zero, a small amount of $H_2O_2$ is consumed, and a small amount of $H_2O$ is produced. This differs from the other two cases, indicating that low-temperature chemistry gradually transitions into intermediate-temperature chemistry. Shortly after this transition, $H_2O_2$ is quickly consumed, and $H_2O$ is rapidly produced, signifying strong intermediate-temperature chemistry and heat release reactions. Consequently, the OSW successfully transitions into ODW. The entire process indicates that the introduced $C_7H_{16}$ vapor quickly participates in reactions and produces heat, facilitating the transition from low-temperature chemistry to intermediate-temperature chemistry and subsequently to heat release reactions, consequently making the transition from OSW to ODW.

### *3.2.4 Effects of on-wedge strips*

It has been demonstrated that introducing n-heptane vapor can enhance the formation of ODW in the combustor. However, the inflow conditions for the ODWE may undergo complex changes, affecting the injection and mixing of liquid fuel. Consequently, the gas-liquid ratio may fluctuate over time and be difficult to control. Therefore, an active control method for ODW formation is necessary. In this section, an on-wedge strip set on the upper wall is used to induce ODW, as proposed by Han et al. [33] in their experiment. To computationally investigate the impact of the strip size, we employ two different sizes (2 mm and 5 mm) of the strip. Other conditions are identical to those used in the benchmark case.

Figure 12 (a) to (c) illustrate the flow field for the case with the on-wedge strip of 5 mm at $t = 1.0$ ms, when the flow field has reached a stable state. The figures reveal that the strip obstructs the flow after the OSW, forming a bow shock wave (BSW) with a relatively large wave angle. This BSW will reflect at the combustor lip, preventing the formation of the SSW. The temperature behind the BSW exceeds 2000 K. The n-heptane droplets quickly diminish post-BSW, and the $C_7H_{16}$ vapor mass fraction also shows a sudden decrease.

This indicates that the n-heptane droplets undergo rapid breakup and evaporation, and the generated vapor swiftly participates in the gaseous reaction. Consequently, this BSW is also a detonation wave.

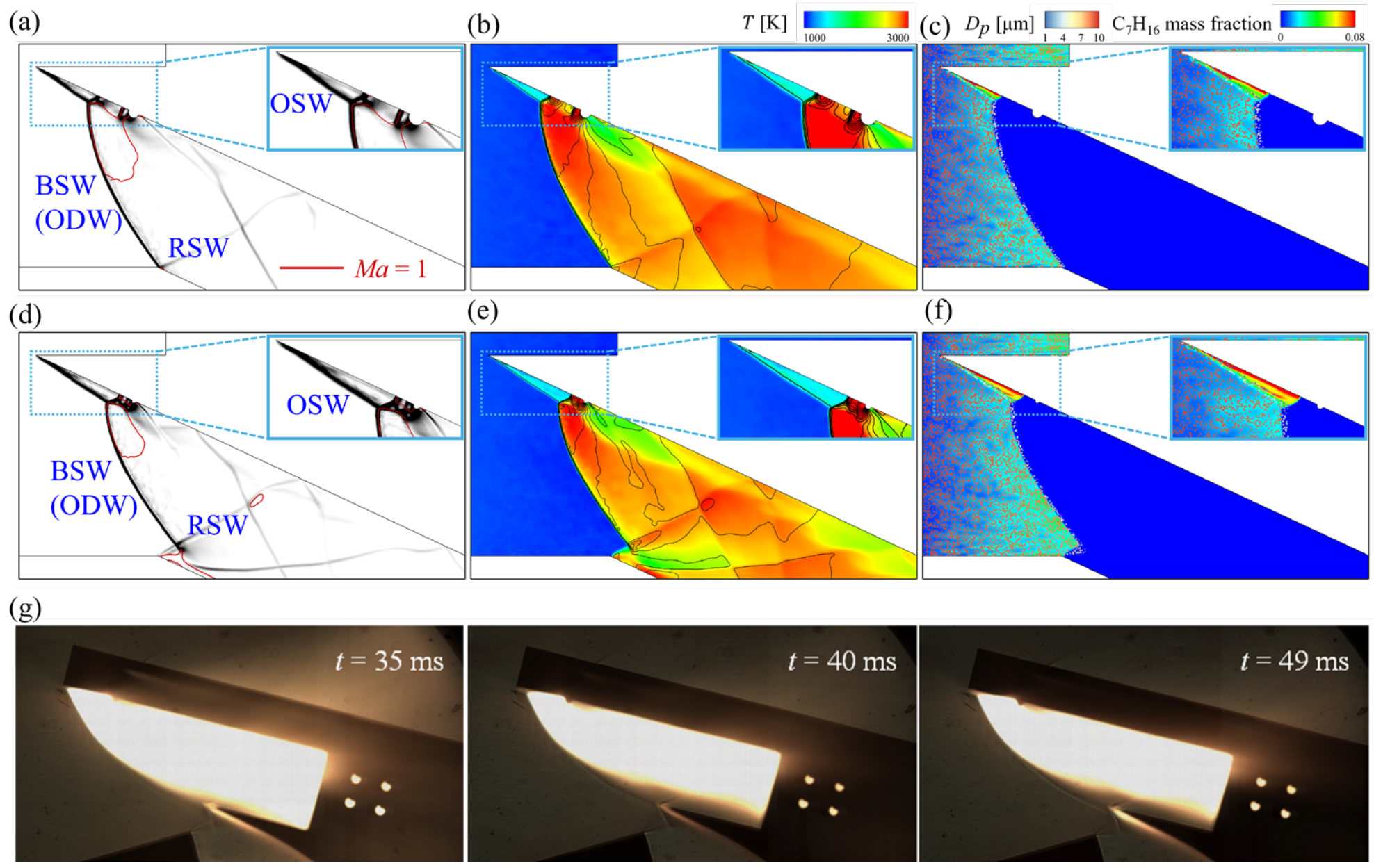

Fig. 12 Flow fields in the combustor with the introduction of the on-wedge strip. (a) Numerical schlieren graph, (b) temperature contour, and (c) C7H16 mass fraction contour with droplet distribution for the case with a 5 mm diameter on-wedge strip. (d) Numerical schlieren graph, (e) temperature contour, and (f) C7H16 mass fraction contour with droplet distribution for the case with a 2 mm diameter on-wedge strip. (g) Colored schlieren graphs from the experiment by Han et al. [33].

For the flow field with a smaller strip of 2 mm, shown in Fig. 12 (d) to (e), the basic flow structures are similar, but the smaller strip delays the appearance of the BSW and the transition from OSW to BSW. Consequently, the BSW reflects after the lip of the combustor, resulting in the formation of a SSW. Despite this, the BSW remains a detonation wave, as indicated by the near-zero $C_7H_{16}$ vapor mass fraction following it. Figure 12 (g) presents the colored schlieren graphs by Han et al. [33] with the introduction of the on-wedge strip with a diameter of 5 mm. Although the details differ between the numerical results due to varying inflow conditions, both the experiment and the simulation reveal the presence of a short OSW and a BSW acting as a detonation wave.

To investigate the formation mechanism with the presence of on-wedge strips from the perspective of chemical kinetics, we show the flow parameters along the streamlines near the upper wall in Fig. 13, with the positions of the OSW, BSW, and MS marked. These streamlines also pass through the same point as Streamline 1. The green and blue curves denote the parameters for cases with 2 mm diameter and 5 mm diameter on-wedge strips, respectively. The trends of the two cases are very similar, so the case with a small on-wedge strip of 2 mm is used as an example for illustration here. The green curves indicate that the OSW leads to an abrupt increase in temperature, resulting in a quick increase in the $C_7H_{16}$ vapor mass fraction post-wave. With the appearance of the $C_7H_{16}$ vapor, $C_7H_{15}$ begins to form, indicating that the initiation reaction occurs shortly after the OSW. Subsequently, KET gradually builds up, and $H_2O_2$ mass fraction continuously increases, indicating the occurrence of low-temperature chemistry. All these processes are similar to the benchmark case shown in red curves until the BSW appears. When the mixture passes through the BSW, the temperature suddenly increases to above 3000 K, $H_2O_2$ and KET are quickly consumed to zero, while the $H_2O$ mass fraction rapidly increases to a steady value. This indicates that intermediate-temperature chemistry dominates the chain branching reactions, leading to significant heat-release reactions and consequently inducing the detonation combustion in the mixture. Unlike the formation mechanism with the introduction of n-heptane vapor, where the vapor leads to a transition from low- to intermediate-temperature chemistry, the BSW directly induces intermediate-temperature chemistry and heat release reactions to facilitate the formation of ODW.

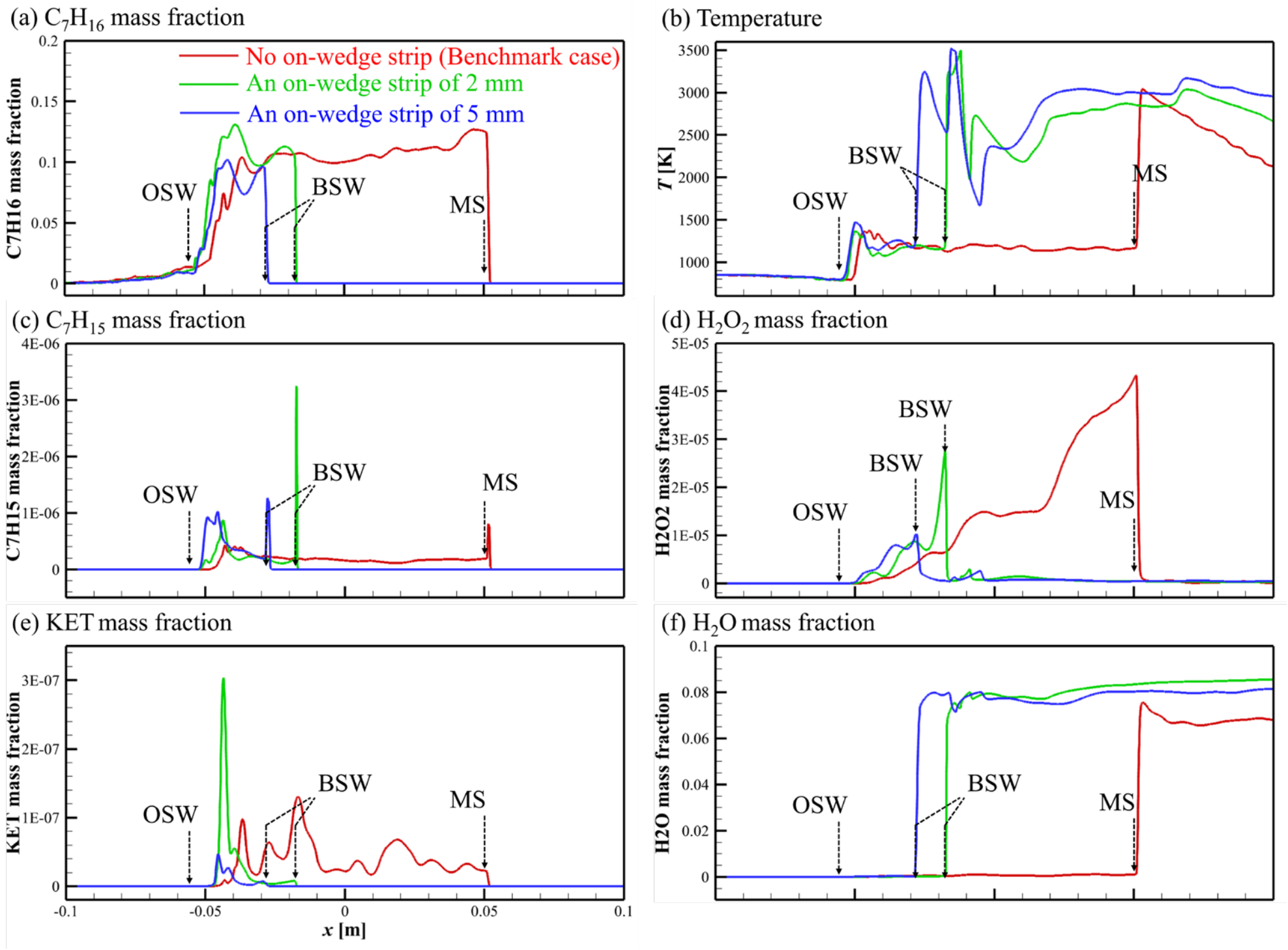

Fig. 13 Flow parameters along streamlines passing through the same point near the upper wall with varying on-wedge strip configurations.

## 4. Conclusions

The ODWE is a hypersonic air-breathing engine distinguished by its rapid combustion processes and high thermal cycle performance, attributed to its operation through ODWs. Liquid hydrocarbon fuels offer significant promise for the ODWE due to their high energy density and convenient storage. However, the utilization of liquid fuels introduces ignition challenges, including extended breakup times, prolonged evaporation, and ignition delays. The formation mechanism of liquid-fueled ODW in the combustor remains inadequately understood up to now. This study employs a computational approach to investigate the ignition of ODWs with n-heptane fuel in a typical two-dimensional ODWE combustor, utilizing a recently developed supersonic two-phase solver based on OpenFOAM. First, the study validates the solver through a comparison of three cases involving wedge-induced gaseous and liquid-fueled ODWs against existing numerical studies. Subsequently, the study explores the formation of ODWs within the typical ODWE combustor. The exploration begins with a benchmark case featuring the pure n-heptane droplets/air inflow, without on-wedge strip, and the Pilch-Erdman breakup model. Based on this benchmark case, this study further examines the effects of various droplet breakup models, the n-heptane vapor proportion, and the on-wedge strips on the ODW formation.

The comparison between the present work and existing studies shows that this solver can accurately simulate gaseous and liquid-fueled ODWs. For cases in the typical combustor, results indicate that the ODW cannot form for the benchmark case, regardless of the breakup models employed. Although the breakup models may influence the flow structure at the intersection point of the OSW and SSW, they will not influence the formation of ODWs.

When the n-heptane vapor is present in the inflow, ODWs can be successfully formed on the upper wall of the combustor. For pure n-heptane/air inflow, the transition from the OSW to the ODW occurs over a relatively shorter length, with ODW occupying most of the combustor. In contrast, for a mixture of n-heptane vapor and droplets, the transition length is extended, resulting in the coexistence of ODW and MS detonation combustion modes. Chemical reaction kinetics analysis reveals that n-heptane vapor can rapidly engage in reactions, generating heat and a variety of intermediate products, facilitating the transition from low-temperature to intermediate-temperature chemical reactions, and ultimately inducing an exothermic reaction that transforms the oblique shock wave into an oblique detonation wave.

The presence of on-wedge strips can also successfully induce the formation of ODWs as demonstrated by previous experiments. The on-wedge strip can form a strong BSW coupled with a thin combustion area, constituting an ODW that occupies most of the combustor, significantly enhances combustion efficiency compared to the benchmark case. Chemical reaction kinetics analysis indicates that the strip-induced BSW significantly elevates the mixture temperature and thereby directly leads to intermediate-temperature chemical reactions and a rapid exothermic process. Consequently, the BSW can directly induce the formation of an ODW.

Although this study computationally explores the influence of droplet breakup models, n-heptane vapor, and on-wedge strips on the formation of ODWs, it does not aim to determine the optimal gas-liquid ratio or the ideal on-wedge strip size, which remain merited subjects for future research. Additionally, the effects of droplet diameter and distribution were not considered and should be addressed for practical applications. Furthermore, the stabilization of ODWs within the combustor, which is significantly influenced by the boundary layer, warrants further investigation.

**Acknowledgements**

This work was supported by the National Natural Science Foundation of China (Grant No. 52176134 and 12172365). The work at the City University of Hong Kong was additionally supported by grants from the Research Grants Council of the Hong Kong Special Administrative Region, China (Project No. CityU 15222421 and CityU 15218820).

## References:

[1] R. DUNLAP, R.L. BREHM, J.A. NICHOLLS, A Preliminary Study of the Application of Steady-State Detonative Combustion to a Reaction Engine, Journal of Jet Propulsion 28 (1958) 451-456.

[2] D.T. Pratt, J.W. Humphrey, D.E. Glenn, Morphology of standing oblique detonation waves, Journal of Propulsion and Power 7 (1991) 837-845.

[3] Y. Fang, Z. Zhang, Z. Hu, X. Deng, Initiation of oblique detonation waves induced by a blunt wedge in stoichiometric hydrogen-air mixtures, Aerospace Science and Technology 92 (2019) 676-684.

[4] K. Kailasanath, Recent Developments in the Research on Pulse Detonation Engines, AIAA Journal 41 (2003) 145-159.

[5] F.K. Lu, H. Fan, D.R. Wilson, Detonation waves induced by a confined wedge, Aerospace Science and Technology 10 (2006) 679-685.

[6] Z. Zhang, K. Ma, W. Zhang, X. Han, Y. Liu, Z. Jiang, Numerical investigation of a Mach 9 oblique detonation engine with fuel pre-injection, Aerospace Science and Technology 105 (2020) 106054.

[7] K. Kailasanath, Review of propulsion applications of detonation waves, AIAA journal 38 (2000) 1698-1708.

[8] E.M. Braun, F.K. Lu, D.R. Wilson, J.A. Camberos, Airbreathing rotating detonation wave engine cycle analysis, Aerospace Science and Technology 27 (2013) 201-208.

[9] C. Li, K. Kailasanath, E.S. Oran, Detonation structures behind oblique shocks, Physics of Fluids 6 (1994) 1600-1611.

[10] C. Viguier, L.F.F.d. Silva, D. Desbordes, B. Deshaies, Onset of oblique detonation waves: Comparison between experimental and numerical results for hydrogen-air mixtures, Symposium (International) on Combustion 26 (1996) 3023-3031.

[11] L.F. Figueria Da Silva, B. Deshaies, Stabilization of an oblique detonation wave by a wedge: a parametric numerical study, Combustion and Flame 121 (2000) 152-166.

[12] M.V. Papalexandris, A numerical study of wedge-induced detonations, Combustion and Flame 120 (2000) 526-538.

[13] H. Teng, H.D. Ng, Z. Jiang, Initiation characteristics of wedge-induced oblique detonation waves in a stoichiometric hydrogen-air mixture, Proceedings of the Combustion Institute 36 (2017) 2735-2742.

[14] Y. Fang, Y. Zhang, X. Deng, H. Teng, Structure of wedge-induced oblique detonation in acetylene-oxygen-argon mixtures, Physics of Fluids 31 (2019).

[15] Y. Zhang, P. Yang, H. Teng, H.D. Ng, C. Wen, Transition Between Different Initiation Structures of Wedge-Induced Oblique Detonations, AIAA Journal 56 (2018) 4016-4023.

[16] B. Bomjan, S. Bhattrai, H. Tang, Characterization of induction and transition methods of oblique detonation waves over dual-angle wedge, Aerospace Science and Technology 82-83 (2018) 394-401.

[17] G. Xiang, X. Li, X. Sun, X. Chen, Investigations on oblique detonations induced by a finite wedge in high altitude, Aerospace Science and Technology 95 (2019) 105451.

[18] Z. Zhang, C. Wen, W. Zhang, Y. Liu, Z. Jiang, A theoretical method for solving shock relations coupled with chemical equilibrium and its applications, Chinese Journal of Aeronautics 35 (2022) 47-62.

[19] J. Verreault, A.J. Higgins, R.A. Stowe, Formation of transverse waves in oblique detonations, Proceedings of the Combustion Institute 34 (2013) 1913-1920.
[20] H. Teng, H.D. Ng, K. Li, C. Luo, Z. Jiang, Evolution of cellular structures on oblique detonation surfaces, Combustion and Flame 162 (2015) 470-477.
[21] B. Parent, J.P. Sislian, J. Schumacher, Numerical Investigation of the Turbulent Mixing Performance of a Cantilevered Ramp Injector, AIAA Journal 40 (2002) 1559-1566.
[22] T.E. Schwartzentruber, J.P. Sislian, B. Parent, Suppression of Premature Ignition in the Premixed Inlet Flow of a Shcramjet, Journal of Propulsion and Power 21 (2005) 87-94.
[23] D.C. Alexander, J.P. Sislian, Computational Study of the Propulsive Characteristics of a Shcramjet Engine, Journal of Propulsion and Power 24 (2008) 34-44.
[24] J.P. Sislian, R.P. Martens, T.E. Schwartzentruber, B. Parent, Numerical Simulation of a Real Shcramjet Flowfield, Journal of Propulsion and Power 22 (2006) 1039-1048.
[25] Z. Zhang, C. Wen, W. Zhang, Y. Liu, Z. Jiang, Formation of stabilized oblique detonation waves in a combustor, Combustion and Flame 223 (2021) 423-436.
[26] Z. Zhang, C. Wen, C. Yuan, Y. Liu, G. Han, C. Wang, Z. Jiang, An experimental study of formation of stabilized oblique detonation waves in a combustor, Combustion and Flame 237 (2022) 111868.
[27] Z. Ren, B. Wang, G. Xiang, L. Zheng, Effect of the multiphase composition in a premixed fuel–air stream on wedge-induced oblique detonation stabilisation, Journal of Fluid Mechanics 846 (2018) 411-427.
[28] Z. Ren, B. Wang, Numerical study on stabilization of wedge-induced oblique detonation waves in premixing kerosene-air mixtures, Aerospace Science and Technology 107 (2020) 106245.
[29] Z. Ren, B. Wang, L. Zheng, Wedge-induced oblique detonation waves in supersonic kerosene-air premixing flows with oscillating pressure, Aerospace Science and Technology 110 (2021) 106472.
[30] Z. Huang, M. Zhao, Y. Xu, G. Li, H. Zhang, Eulerian-Lagrangian modelling of detonative combustion in two-phase gas-droplet mixtures with OpenFOAM: Validations and verifications, Fuel 286 (2021) 119402.
[31] H. Guo, Q. Meng, Y. Xu, H. Zhang, Autoignition of two-phase n-heptane/air mixtures behind an oblique shock: Insights into spray oblique detonation initiation, Combustion and Flame 256 (2023) 112992.
[32] H. Teng, C. Tian, P. Yang, M. Zhao, Effect of droplet diameter on oblique detonations with partially pre-vaporized n–heptane sprays, Combustion and Flame 258 (2023) 113062.
[33] X. Han, Y. Liu, Z. Zhang, W. Zhang, C. Yuan, G. Han, Z. Jiang, Experimental demonstration of forced initiation of kerosene oblique detonation by an on-wedge trip in an ODE model, Combustion and Flame 258 (2023) 113102.
[34] S. Liu, J.C. Hewson, J.H. Chen, H. Pitsch, Effects of strain rate on high-pressure nonpremixed n-heptane autoignition in counterflow, Combustion and flame 137 (2004) 320-339.
[35] C. Qi, P. Dai, H. Yu, Z. Chen, Different modes of reaction front propagation in n-heptane/air mixture with concentration non-uniformity, Proceedings of the Combustion Institute 36 (2017) 3633-3641.
[36] Q. Meng, M. Zhao, Y. Xu, L. Zhang, H. Zhang, Structure and dynamics of spray

detonation in n-heptane droplet/vapor/air mixtures, Combustion and Flame 249 (2023) 112603.
[37] B. Abramzon, W.A. Sirignano, Droplet vaporization model for spray combustion calculations, International journal of heat and mass transfer 32 (1989) 1605-1618.
[38]
R.H. Perry, D.W. Green, J.O. Maloney. Perry's Chemical Engineers' Handbook. In: editor^editors. 2007. p.
[39] W.E. Ranz, Evaporation from drops: Part I, Chemical engineering progress 48 (1952) 141.
[40] C.T. Crowe, M.P. Sharma, D.E. Stock, The Particle-Source-In Cell (PSI-CELL) Model for Gas-Droplet Flows, Journal of Fluids Engineering-transactions of The Asme 99 (1977) 325-332.
[41] P.J. O'Rourke, A.A. Amsden, The TAB method for numerical calculation of spray droplet breakup, Report No. 0148-7191, SAE technical paper, 1987.
[42] M. Pilch, C. Erdman, Use of breakup time data and velocity history data to predict the maximum size of stable fragments for acceleration-induced breakup of a liquid drop, International journal of multiphase flow 13 (1987) 741-757.
[43] R. Reitz, Modeling atomization processes in high-pressure vaporizing sprays, Atomisation and Spray technology 3 (1987) 309-337.
[44] W. Wang, M. Yang, Z. Hu, P. Zhang, A Dynamic Droplet Breakup Model for Eulerian-Lagrangian Simulation of Liquid-fueled Detonation, Aerospace Science and Technology, doi:https://doi.org/10.1016/j.ast.2024.109271(2024) 109271.
[45] S. Sharma, A.P. Singh, S.S. Rao, A. Kumar, S. Basu, Shock induced aerobreakup of a droplet, Journal of Fluid Mechanics 929 (2021) A27.
[46] Y. Zhang, Y. Fang, H.D. Ng, H. Teng, Numerical investigation on the initiation of oblique detonation waves in stoichiometric acetylene–oxygen mixtures with high argon dilution, Combustion and Flame, (2019).
[47] C.J. Greenshields, H.G. Weller, L. Gasparini, J.M. Reese, Implementation of semi-discrete, non-staggered central schemes in a colocated, polyhedral, finite volume framework, for high-speed viscous flows, International journal for numerical methods in fluids 63 (2010) 1-21.
[48] A. Kurganov, S. Noelle, G. Petrova, Semidiscrete central-upwind schemes for hyperbolic conservation laws and Hamilton--Jacobi equations, SIAM Journal on Scientific Computing 23 (2001) 707-740.
[49] C. Tian, H. Teng, B. Shi, P. Yang, K. Wang, M. Zhao, Propagation instabilities of the oblique detonation wave in partially prevaporized n-heptane sprays, Journal of Fluid Mechanics 984 (2024) A16.
[50] H. Teng, C. Tian, Y. Zhang, L. Zhou, H.D. Ng, Morphology of oblique detonation waves in a stoichiometric hydrogen–air mixture, Journal of Fluid Mechanics 913 (2021) A1.
[51] J. Li, Z. Zhao, A. Kazakov, M. Chaos, F.L. Dryer, J.J. Scire Jr, A comprehensive kinetic mechanism for CO, CH2O, and CH3OH combustion, International Journal of Chemical Kinetics 39 (2007) 109-136.
[52] M.P. Burke, M. Chaos, Y. Ju, F.L. Dryer, S.J. Klippenstein, Comprehensive H2/O2 kinetic model for high-pressure combustion, International Journal of Chemical Kinetics 44 (2012) 444-474.